\documentclass[aip,jcp,preprint,longbibliography,floatfix]{revtex4-2}
\usepackage{graphicx}
\usepackage{amsmath, amssymb}
\usepackage{color}
\usepackage{ulem}
\usepackage{hyperref}
\def\eqref#1{(\ref{#1})}
\def\angb#1{\left<#1\right>}
\def\bm#1{\mbox{\boldmath $#1$}}
\def\ptl{\partial}
\def\ie{\textit{i.e.}, }
\def\eg{\textit{e.g.}, }
\def\mrd{\mathrm{d}}
\def\kB{k_\mathrm{B}}
\def\Hamil{\mathcal{H}}

\def\xpw{x^\mathrm{pw}}
\def\xpwa{\xpw_\alpha}
\def\xpwb{\xpw_\beta}
\def\xpwp{x^{\mathrm{pw}\prime}}
\def\xpcl{x^\mathrm{pc}_\mathrm{left}}
\def\xpcr{x^\mathrm{pc}_\mathrm{right}}
\def\VCV{V^\mathrm{CV}}
\def\Wpc{\Delta W_\mathrm{pc}}
\def\Wiso{W_\mathrm{iso}}
\def\pset{p_\mathrm{set}}
\def\Wdet{W_\mathrm{det}}
\def\Wsep{W_\mathrm{sep}}
\def\WisoDS{\Wiso^\mathrm{DS}}
\def\WisoPW{\Wiso^\mathrm{PW}}
\def\Wsl{W_\mathrm{SL}}
\def\Wsv{W_\mathrm{SV}}
\def\glv{\gamma_\mathrm{LV}}
\def\gsl{\gamma_\mathrm{SL}}

\def\gs0{\gamma_\mathrm{S0}}
\def\gl0{\gamma_\mathrm{L0}}
\def\gv0{\gamma_\mathrm{V0}}
\def\gab{\gamma_{\alpha \beta}}

\def\lambdap{\lambda^{\prime}}
\begin{document}
\title{
Mechanical and thermodynamic routes to the liquid-liquid interfacial tension and mixing free energy
by molecular dynamics
}
\newcommand{\osaka}{Department of Mechanical Engineering, Osaka University, 2-1 Yamadaoka, Suita 565-0871, Japan}
\newcommand{\tuswater}{Water Frontier Research Center (WaTUS),
	Research Institute for Science \& Technology,
	Tokyo University of Science,
	1-3 Kagurazaka, Shinjuku-ku, Tokyo, 162-8601, Japan}
\newcommand{\osakamu}{Department of Mechanical Engineering, 
	Osaka Metropolitan University, 3-3-138 Sugimoto, Sumiyoshi, Osaka 558-8585, Japan}
\newcommand{\tohoku}{Department of Mechanical Systems Engineering, Tohoku University, 6-6 Aramaki Aoba-ku, Sendai 980-8579, Japan}
\newcommand{\brunel}{Department of Mechanical and Aerospace Engineering, Brunel University London, Uxbridge, UB8 3PH United Kingdom}
\newcommand{\ilm}{Universit{\'e} Claude Bernard Lyon 1, CNRS, Institut Lumière Matière, UMR5306, F69100 Villeurbanne, France}
\author{Rei Ogawa}%
\affiliation{\osaka}
\author{Hiroki Kusudo}%
\email{kusudo@tohoku.ac.jp}
\affiliation{\tohoku}
\author{Takeshi Omori}%
\email{t.omori@omu.ac.jp}
\affiliation{\osakamu}
\author{Edward R. Smith}%
\email{Edward.Smith@brunel.ac.uk}
\affiliation{\brunel}
\author{Laurent Joly}%
\email{laurent.joly@univ-lyon1.fr}
\affiliation{\ilm}
\author{Samy Merabia}%
\email{samy.merabia@univ-lyon1.fr}
\affiliation{\ilm}
\author{Yasutaka Yamaguchi}
\email{yamaguchi@mech.eng.osaka-u.ac.jp}
\affiliation{\osaka}
\affiliation{\tuswater}
\date{\today}

\begin{abstract}
In this study, we carried out equilibrium molecular dynamics (EMD) simulations of the liquid-liquid interface between two different Lennard-Jones components with varying miscibility, where we examined the relation between the interfacial tension and isolation free energy using both a mechanical and thermodynamic approach.
Using the mechanical approach, we obtained a stress distribution around a quasi-one-dimensional (1D) EMD systems with a flat LL interface. From the stress distribution, we calculated the liquid-liquid interfacial tension based on Bakker's equation, which uses the stress anisotropy around the interface, and measures how it varies with miscibility. The second approach uses thermodynamic integration by enforcing quasi-static isolation of the two liquids to calculate the free energy. This uses the same EMD systems as the mechanical approach, with both extended dry-surface and phantom-wall (PW) schemes applied. 
When the two components were immiscible, the interfacial tension and isolation free energy were in good agreement, provided all kinetic and interaction contributions were included in the stress. When the components were miscible, the values were significantly different. From the result of PW for the case of completely mixed liquids, the difference was attributed to the additional free energy required to separate the binary mixture into single components against the osmotic pressure prior to the complete detachment of the two components, \ie the free energy of mixing.
\end{abstract}
\maketitle 
\section{Introduction}
\label{sec:intro}
When two immiscible liquids, such as water and oil, coexist, an interface is typically formed between them, referred to as a liquid-liquid (LL) interface, where a LL interfacial tension arises.
	Such LL interface that can be found in emulsions, which are ubiquitous in our daily life, \eg food, drink and cosmetics, so that measuring the LL interfacial free energy and/or tension as a macroscopic property is one of the important tasks for the control of the properties of the mixture. 
\par
From a microscopic point of view, Kirkwood  and Buff~\cite{kirkwood_statistical_1935,kirkwood_statistical_1951}  formulated 
expressions for the chemical potentials of the 
components of gas mixtures and liquid solutions 
based on statistical mechanics, and they also provided a general theoretical framework to describe surface tension.~\cite{Kirkwood1949}
Related to this point, 
with respect to a liquid-vapor (LV) or liquid-gas (LG) interface, Bakker's equation~\cite{Bakker1928}, based on macroscopic thermodynamics, was known before the formulation from statistical mechanics mentioned above. This equation relates the macroscopic LV or LG interfacial tension to the 
anisotropy of the microscopic local stress near the interface. 
For a 
flat LV interface normal to the $x$-axis, Bakker's equation is written as
\begin{equation}
	\gamma_\mathrm{LV}=
	\int_{x^{\mathrm{blk}}_\mathrm{V}}^{x^{\mathrm{blk}}_\mathrm{L}}
	\left[\tau_{yy}(x)-\tau_{xx} \right] \mrd x
	=
	\int_{x^{\mathrm{blk}}_\mathrm{V}}^{x^{\mathrm{blk}}_\mathrm{L}}
	\tau_{yy}(x) \mrd x 
	-
	\int_{x^{\mathrm{blk}}_\mathrm{V}}^{x^{\mathrm{blk}}_\mathrm{L}}
	\tau_{xx} \mrd x 
	\label{eq:Bakker's eq}
\end{equation}
where $\glv$ is the LV interfacial tension, and $\tau_{yy}$ $(=\tau_{zz})$ and  $\tau_{xx}$ are the diagonal stress components tangential and normal to the interface, respectively. 
From the local force balance in the direction normal to the interface, $\tau_{xx}$ is constant over the entire region, which is equal to the isotropic pressure in the bulk with its sign inverted. 
By integrating the stress difference $\tau_{yy}(x)-\tau_{xx}$ existing only around the interface, \ie by taking $x^{\mathrm{blk}} _{\mathrm{L}}$ and $x^{\mathrm{blk}} _{\mathrm{V}}$ in Eq.~\eqref{eq:Bakker's eq} as the bulk positions of the liquid and gas phases, respectively, the LV interfacial tension $\glv$ is obtained. 
\par
At present, molecular dynamics (MD) simulation is a powerful tool to investigate LV or LG interfaces as well as LL mixtures composed of various kinds of molecular pairs in silico.
A quasi-one-dimensional (1D) MD system can be easily simulated, \eg one with LV coexistence can be constructed by confining a single molecular component in a constant-volume simulation cell with PBCs, at a temperature between the triple point and the critical point, with setting proper number of molecules in the cell. 
In such a MD system, the two integrals in the right-most hand-side of Bakker's equation~\eqref{eq:Bakker's eq} can be obtained,~\cite{Allen1989} and thus Eq.~\eqref{eq:Bakker's eq} is widely used to calculate the surface tension as a standard approach in MD. This is partly because only the integral of each principal stress component in the whole system is used, which can be easily calculated in such a quasi-1D system. Nevertheless, it should also be noted that proper calculation methods 
of the local stress called the volume average (VA) or the method 
of plane (MoP) are needed
in order to accurately obtain the local stress, so that the stress 
field can satisfy the mechanical requirements, \eg 
satisfying $\tau_{xx}=\mathrm{const.}$ over the entire region 
mentioned above.~\cite{Todd1995,Shi2023, Nishida2014, Yamaguchi2019} 
\par
Regarding the LL interface, in the early stage of MD development, \citet{Hayoun1988} simulated a quasi-1D
system with an interface between two Lennard-Jones (LJ) liquids and showed the density and pressure distribution, and \citet{benjamin_molecular_1997} wrote a review 
article of MD studies regarding the  structure and dynamics at the LL interface. At present, the LL interfacial tension is also calculated using the integrated form of Bakker's equation similar to  Eq.~\eqref{eq:Bakker's eq} as a definition,~\cite{feria_molecular_2022} 
although the stress definition for mixture liquids is not so straightforward as that for single-component liquid.\cite{Sega2016,Hantal2020}
%
Furthermore, the authors have successfully extended Bakker's equation~\eqref{eq:Bakker's eq} to flat solid-liquid (SL) and 
solid-vapor (SV) interfaces. In this framework, through a careful
choice of the SL and SV interface positions based on a mechanical 
force balance, the microscopic interpretation of Young's equation 
was clarified as the force balance exerted on the fluid particles 
in a finite region surrounding the contact line. 
	It should be noted that the microscopic stress calculated by the 
	MoP was evaluated only by including the interaction forces between 
	fluid particles while the force from the solid particles on the 
	fluid particles were treated as an external force. 
Such a treatment of interfacial tensions including surface tension
based on a mechanical force balance is called the mechanical route.~\cite{Leroy2009,Leroy2015,Kanduc2017,Yamaguchi2019,Russo2019}
\par
Besides the mechanical route, the SL and SV 
interfacial tensions have been calculated as the free energy per unit area of the interface
based on thermodynamic integration methods.~\cite{Surblys2018, Yamaguchi2019, Bistafa2021, Shintaku2024} In this route, the 
solid-fluid (SF) work of adhesion was calculated by 
quasi-statically isolating the solid and fluid sandwiching 
a depletion layer by using a virtual wall interacting only 
with the fluid, or by gradually reducing the SF interaction 
strength so that the solid and fluid did not interact with 
each other. 
More concretely, the quasi-static SL 
isolation process was carried out 
keeping the number of particles $N$, pressure $p$ and 
temperature $T$, \ie under constant $NpT$ condition, 
and the SL work of adhesion $\Wsl$ was obtained from 
the change of Gibbs free energy $G$ of the system given by
\begin{equation}
	\Wsl \equiv \frac{\Delta G}{A}
	= 
	\gs0 + \gl0 - \gsl,
\end{equation}
where $A$ is the interfacial area and $\gamma$ is
the interfacial free energy per area with corresponding
interface denoted by the subscript (`S':solid, `L':liquid
and `0':depletion layer). Note that $\Delta G$ excluded 
the work exerted on the environment.~\cite{Yamaguchi2019} 
The solid-vapor 
work of adhesion $\Wsv$ was evaluated as well by similar schemes.
%
As a result, it was shown that the mechanically-obtained 
SL and SV interfacial tensions and thermodynamically-obtained
works of adhesion agreed well for a LJ fluid on a simple crystal surface. 
%
\par
	In this study, we investigate the possibility to extend 
	the mechanical and thermodynamic routes to the LL interface
	with a focus on the following points: 
	1) whether Bakker's equation as a mechanical route can 
	be extended to LL interface, and how to implement the 
	stress calculation in a system with 
	two different liquid components, and 
	2) whether the mechanical route corresponds to the 
	thermodynamic route for the LL interface. Related to 
	the second point, we tested two methods for the 
	thermodynamic route to validate the results. In 
	addition, we show that one of the thermodynamic 
	methods can provide a clear insight into the 
	\emph{isolation} free energy, 
	\ie the free energy needed to separate the mixture into single components --which includes the free energy of mixing and the change in interfacial free energy, and also gives access to the osmotic pressure of the liquid mixtures. %
\section{Method}
\label{sec:method}
\subsection{Simulation setup}
\label{subsec:simulation}
\begin{figure}[t]
	\begin{center}
		\includegraphics[scale=0.4]{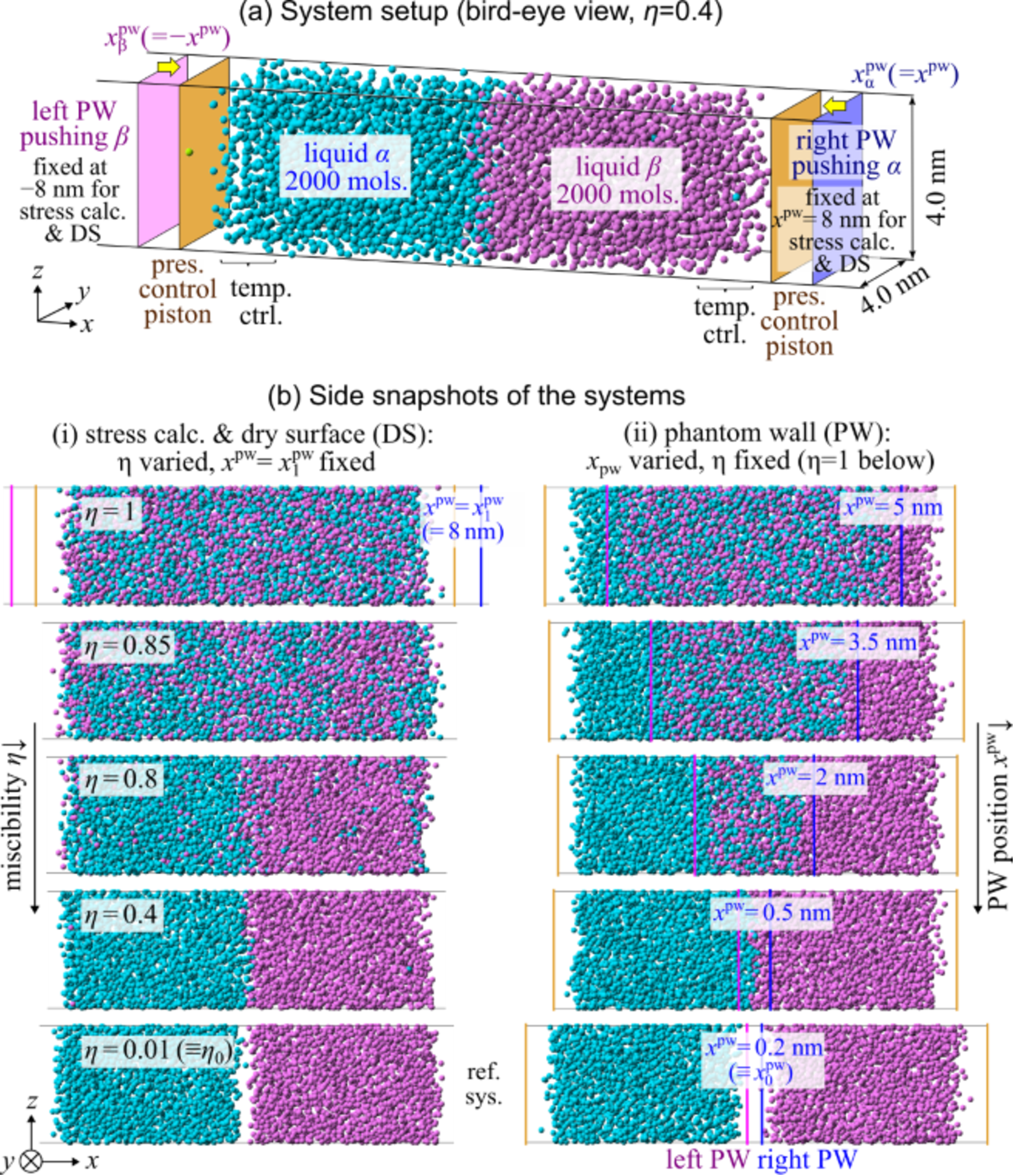}
	\end{center}
	\caption{
		\label{fig:system}		
		(a) Setup of the equilibrium simulation system with a liquid-liquid 
		interface controlled by a miscibility parameter $\eta$. 
		A pressure control piston and a semipermeable phantom wall (PW) 
		were located on each end of the system in the $x$-direction 
		(left and right).
		(b) Side snapshots of the equilibrium systems for the free-energy calculation by the thermodynamic integration using
		the (i) extended dry surface (DS), and (ii) extended 
		PW methods.
		For the extended DS method (i), 
		the miscibility parameter $\eta$ was varied while keeping 
		the PWs at $\xpw_{1}$ and $-\xpw_{1}$ far from the 
		liquid; thus, the PWs as well as the PCs are shown only on the top panel of (i). 
		For the PW method (ii), the PW positions $\xpw$ and $-\xpw$ 
		were varied while keeping $\eta$ unchanged.
	}
\end{figure}
Figure~\ref{fig:system} shows the quasi-1D 
equilibrium MD (EMD) system studied. 
Two kinds of  LJ particles denoted by $\alpha$ and $\beta$ having the same mass of $m=6.634\times 10^{-26}$~kg equal to the mass of an argon atom of 40~amu, were confined between two walls on the left and right ends 
both parallel to the $yz$-plane shown in light brown. 
The numbers of particles $\alpha$ and $\beta$ were both set equal 
to 2,000.  The interaction potentials between the same components $\phi_{\alpha\alpha}$ and $\phi_{\beta\beta}$ were expressed by
\begin{equation}
	\label{eq:lj_potential}
	\phi_{\alpha \alpha}{\left(r_{ij}\right)}
	=
	\phi_{\beta \beta}{\left(r_{ij}\right)}
	= 4\varepsilon\left[  
	\left(\frac{\sigma}{r_{ij}}\right)^{12} - \left(\frac{\sigma}{r_{ij}}\right)^{6}
	\right],
\end{equation}
where $r_{ij}$ denotes the distance between particles $i$ and $j$ of the same component $\alpha$ or $\beta$, and $\sigma$ and $\varepsilon$ were the LJ length and energy potential parameters, which we set as $\sigma=0.34~\mathrm{nm}$ and $\varepsilon=1.67\times10^{-21}~\mathrm{J}$.
%
%
%
The interaction potential between different components $\phi_{\alpha\beta}(r_{ij})$ was 
expressed by multiplying the interaction potential in Eq.~\eqref{eq:lj_potential} by $\eta$ 
as:
\begin{equation}
	\label{eq:lj_potential_eta}
	\phi_{\alpha \beta}{\left(r_{ij}\right)}
	= \eta \phi_{\alpha\alpha}{\left(r_{ij}\right)}
\end{equation}
where $\eta$ was set between 0.01 and 1 as a variable parameter.
\par
Periodic boundary conditions were adopted in  $y$- and 
$z$-directions, and the cell size in these directions 
$L_{y}$ and $L_{z}$ were both 4~nm.
In the following, the system $yz$ cross sectional area is denoted $A=L_{y}L_{z}$.
In addition, we located a pressure control (PC) wall 
and a semipermeable phantom wall (PW) on each end of the system 
in the $x$-direction 
(left and right), all of which were parallel to the $yz$-plane and 
interacted with the fluid particles as a unique function of the 
distance given by
\begin{equation}
	\phi_\mathrm{fw}(x'_{i})=
	\frac{4}{5} \pi \varepsilon_\mathrm{fw} \rho_\mathrm{s} x^{\prime\,2}_{i}
	\left( \frac{\sigma_\mathrm{fw}}{x^{\prime}_{i}}\right)^{12},
\end{equation}
where $x^{\prime}_{i}$ is the distance between particle $i$ at $x_{i}$
and corresponding wall. 
This potential field corresponds to a mean potential field formed by a single wall layer of uniformly distributed solid particles with an area
number density $\rho_\mathrm{s}=(3.61)^{2}$~nm$^{-2}$, which interact with the fluid particles through
the LJ potential only with repulsive term (Eq.~\eqref{eq:lj_potential} without $(\sigma/r_{ij})^{6}$ in the RHS) with the energy and length parameters being $\varepsilon_\mathrm{fw}=1.29\times10^{-21}$~J 
and $\sigma_\mathrm{fw}=0.345$~nm, respectively. 
The presented results are not sensitive to the choice of parameter values, which are set the same as in our previous studies,~\cite{Yamaguchi2019,Shintaku2024} as long as the interaction is repulsive and short-ranged.
\par
For the PCs on the left and right at $\xpcl$
and $\xpcr$, respectively shown in light brown in Fig.~\ref{fig:system}~(a), 
we used
\begin{gather}
\nonumber
\phi_\mathrm{fluid-pc}(x'_{i})=\phi_\mathrm{fw}(x'_{i})
\\
x'_{i} = x_{i} - \xpcl, \quad
x'_{i} = \xpcr - x_{i}\quad
(i\in \alpha, \beta).
\label{eq:potential_pcs}
\end{gather}
By adjusting the positions $\xpcl$ and $\xpcr$ of the walls
as pistons, the system pressure was maintained constant 
at $p_\mathrm{set} \approx 1~\mathrm{MPa}$. 
\par
On the other hand, for the PWs on the left and right at $\xpwa$
and $\xpwb$, respectively, we applied
\begin{equation}
\phi_{\mathrm{fluid-pw}\alpha}(x'_{i})=\phi_\mathrm{fw}(x'_{i}),
\quad
x'_{i} = \xpwa - x_{i}\quad (i\in \alpha)
\label{eq:potential_pwa}
\end{equation}
and
\begin{equation}
\phi_{\mathrm{fluid-pw}\beta}(x'_{i})=\phi_\mathrm{fw}(x'_{i}),
\quad
x'_{i} = x_{i}-\xpwb \quad (i\in \beta)
\label{eq:potential_pwb}
\end{equation}
with setting
\begin{equation}
\xpwa = \xpw, \quad \mathrm{and}\quad \xpwb = -\xpw.
\label{eq:xpwb=-xpwa}
\end{equation}
With this setting, the PWs interacted only with $\alpha$ or 
$\beta$ and worked as semipermeable membranes located 
symmetric to the $yz$-plane.
\par
The system temperature $T$ was controlled at 85~K by applying a velocity 
rescaling thermostat to the fluid particles located at less than $1.5~\mathrm{nm}$ from the potential 
walls, only for the velocity components in the $y$- and $z$-directions. 
These thermostat regions were sufficiently away from the interface 
and no direct thermostating was applied to the region near the interface 
so that this thermostat had no effects on the presented results.
\par
For the calculation of the stress distribution as a mechanical route 
and for the dry surface (DS) method as a thermodynamic route 
shown in Fig.~\ref{fig:system}~(b-i), the symmetric PW 
position was fixed at $\xpw (= \xpwa=-\xpwb)=8$~nm, at 
which the PWs were sufficiently far away from the liquid so that they 
did not interact with the fluids. We denote this value of $\xpw=8$~nm
as $\xpw_{1}$ hereafter.
With this setting, a quasi-1D 
system under constant $NpT$ was obtained as an equilibrium state 
for each $\eta$ value. The two liquids were completely mixed without 
interface at $\eta=1$ because both liquids are identical. By decreasing 
$\eta$, the two liquids were separated at $\eta=0.85$ and formed 
a flat LL interface, and at $\eta=0.01$, the two liquids were 
isolated with a vacant region between two liquids in this case.
This value of $\eta=0.01$ is denoted as $\eta_{0}$ hereafter.
\par
For the extended phantom wall (PW) method in Fig.~\ref{fig:system}~(b-ii)
as another thermodynamic route described below, the PW position 
$\xpw$ was changed with keeping $\eta$ unchanged. With the decrease of 
$\xpw$, the two liquids were separated by the semipermeable PWs,
and they were isolated at $\xpw = 0.2$~nm in this case.
This value of $\xpw=0.2$~nm is denoted as $\xpw_{0}$ hereafter.
\subsection{Mechanical route}
\label{subsec:mechanical route}
Here, we describe how Bakker's equation is used to calculate the LL interfacial tension from the stress distribution as a mechanical route. 
Equation~\eqref{eq:Bakker's eq} extended to the LL interface 
can be written as 
\begin{equation}	
\gamma_{\alpha\beta}=\int_{x^{\mathrm{blk}} _\alpha}^{x^{\mathrm{blk}} _\beta}\left[\tau_{yy}(x)-\tau_{xx} \right]\mrd x,
\label{eq:ex_Bakker's eq}
\end{equation}
where $\gab$ is the LL interfacial tension, and $x^\mathrm{blk} _{\alpha}$ and $x^\mathrm{blk}_{\beta} (>x^\mathrm{blk} _{\alpha})$ are the bulk positions of the liquid phases $\alpha$ and $\beta$, respectively, 
where the stress is isotropic.  
As mentioned in the introduction, defining stress for a mixture of liquids is more complex than for a single-component liquid. 
In general, the stress tensor $\bm{\tau}$ is expressed by
the sum of a kinetic term $\bm{\tau}^\mathrm{kin}$ and an interaction 
term $\bm{\tau}^\mathrm{int}$ as
\begin{equation}
\bm{\tau} = 
\bm{\tau}^\mathrm{kin} + 
\bm{\tau}^\mathrm{int}, 
\label{eq:tau=kin+int}
\end{equation}
and it is not straightforward to determine how to incorporate the kinetic contribution from each component and interaction contribution from intermolecular potential between fluid particles of different components into the local stress calculation. 
In this study, the fluid was composed of two monoatomic components $\alpha$ and $\beta$, and we assumed that all contributions from the two 
components should be included in each term. 
At first, the kinetic energy term $\bm{\tau}^\mathrm{kin}$ can 
be written by
\begin{align}
\bm{\tau}^\mathrm{kin}
=
\bm{\tau}^{\mathrm{kin},\alpha}
+
\bm{\tau}^{\mathrm{kin},\beta},
\label{eq:tau^kin_ab}
\end{align}
where the superscript `kin,$\alpha$' for instance, denotes 
the contribution from $\alpha$ particles to the kinetic term. 
On the other hand, the interaction term $\bm{\tau}^\mathrm{int}$ is written by
\begin{equation}
\bm{\tau}^\mathrm{int}
=
\bm{\tau}^{\mathrm{int},\alpha\alpha}
+
\bm{\tau}^{\mathrm{int},\beta\beta}
+
\bm{\tau}^{\mathrm{int},\alpha\beta},
\label{eq:tau^int_ab}
\end{equation}
where the superscript `int,$\alpha\beta$' denotes, for instance, 
the contribution from the intermolecular interaction between $\alpha$
and $\beta$ particles to the interaction term.
Since it was not obvious whether all terms on the right-hand side (RHS) of each equation should be included as in Eqs.~\eqref{eq:tau^kin_ab} and \eqref{eq:tau^int_ab}, we verified that by checking if the stress definition satisfied the local mechanical balance 
\begin{equation}
\nabla \cdot \bm{\tau} = \bm{0}
\end{equation}
in equilibrium systems at an arbitrary point in the absence of the external field,~\cite{Shi2023} \ie except near the PCs and PWs. 
In practice, we calculated the stress distribution by the volume average 
(VA) method,~\cite{Shi2023,Nishida2014} and tested if
\begin{equation}
\tau_{xx} = \mathrm{const.}
\label{eq:tau=const}
\end{equation}
was satisfied in the present quasi-1D system for all possible combinations
of the terms in the RHS of Eqs.~\eqref{eq:tau^kin_ab} and \eqref{eq:tau^int_ab}.
\par
We applied the VA approach~\cite{Shi2023,Nishida2014} to calculate the stress in local flat regions with a thickness $\delta x$ and a volume 
$\VCV$ of $A\delta x$, 
where the subscript `CV' stands for `control volume.'
The two terms of the kinetic contribution in the RHS of Eq.~\eqref{eq:tau^kin_ab} were expressed by
\begin{align}
\bm{\tau}^{\mathrm{kin},\xi}=
-
\frac{1}{\VCV}
\left<
\sum_{
			i \in \xi
		}
		^{N_\xi} m_i \bm{v}_i \bm{v}_i \vartheta_i
		\right>,
		\quad
		(\xi = \alpha, \beta),
		\label{eq:tau^kin_ab_VA}
	\end{align}
	where $\bm{r}_{i}$ and $\bm{v}_{i}$ are the position and velocity 
	vectors of $i$-th $\xi$ particle, and 
	$\vartheta_i$ is a function which is one if the particle is 
	in the local volume or zero otherwise, and the summation is taken over all $N_{\xi}$ particles in the system. The angle brackets 
	denote the ensemble average; in practice, this ensemble 
	average is usually substituted by the time average in 
	steady-state MD systems including EMD  
	ones,~\cite{Kusudo2021, Kusudo2023} 
	and a moving time average in non-equilibrium MD.~\citep{Todd_Daivis_book}
	%
%
On the other hand, for a simple two-body interaction between the particles, the three terms of the interaction contribution in the RHS of Eq.~\eqref{eq:tau^int_ab} are separated into the following
\begin{equation}
	\bm{\tau}^{\mathrm{int},\xi\zeta} 
	= \left\{
	\begin{array}{ll}
		-\displaystyle \frac{1}{\VCV}\left<
		\underset{i,j(>i)\in \xi}{\sum^{N_{\xi}-1}\sum^{N_{\xi}}}
		w_{ij}^\mathrm{CV} \bm{r}_{ij} \otimes \bm{f}_{ij}
		\right>,
		& (\xi=\zeta=\alpha,\beta)
		\\
		-\displaystyle \frac{1}{\VCV}\left<
		\sum_{i\in \xi}^{N_{\xi}}\sum_{j\in \zeta}^{N_{\zeta}}
		w_{ij}^\mathrm{CV} \bm{r}_{ij} \otimes \bm{f}_{ij}
		\right>,
		& (\xi=\alpha,\ \zeta=\beta)
	\end{array}
	\right.,
	\label{eq:tau^int_ab_VA}
\end{equation}
where $\bm{r}_{ij}=\bm{r}_{j}-\bm{r}_{i}$ and $\bm{f}_{ij}$ are the relative position
vector and force exerted from particle $i$ to $j$, 
whereas $w_{ij}^\mathrm{CV}$ denotes the weighting function 
given as the length fraction of the straight line segment connecting 
particle $i$ and $j$ in the CV. 
A mathematically proper expression for the Cartesian coordinate system is given in Ref.~\citenum{Shi2023}.
\par
Indeed as naturally expected from equilibrium momentum balance, 
Eq.~\eqref{eq:tau=const} was satisfied for all systems with various
$\eta$ values only when all terms in Eqs.~\eqref{eq:tau^kin_ab_VA} and \eqref{eq:tau^int_ab_VA} were included. This point will be further
discussed in Sec.~\ref{sec:resdis}.
%
Based on this result, the interfacial tension $\gab$ was obtained by Eq.~\eqref{eq:ex_Bakker's eq} using the stress distributions obtained by the VA approach.
\subsection{Thermodynamic route}
\label{subsec:thermodynamic route}
As a thermodynamic approach, Leroy~\textit{et al.}~\cite{Leroy2009, Leroy2010, Leroy2015} proposed to calculate the SL and SV interfacial tensions using the thermodynamic integration (TI) method.
%
Generally, TI is a method to calculate the free energy difference between two different equilibrium systems by connecting them by a thermodynamically reversible quasi-static route.~\cite{Frenkel2001,Davidchack2003,diPasquale2022} 
For example, one introduces a coupling parameter $\lambda$ into a certain constant which represents the state of a system under $NpT$ ensemble, and denotes the Hamiltonian of the system by $\Hamil(\bm{\Gamma}, \lambda)$ using $\lambda$ and the phase variable $\bm{\Gamma}$ composed of the positions and momenta of all constituent particles.
Let $G(\lambda)$ be defined by the Gibbs free energy of the system
as a function of $\lambda$.
Then, its difference between a target system at $\lambda = \lambda_{1}$ 
and a reference system at $\lambda = \lambda_{0}$ is obtained as follows:
\begin{align}
	G(\lambda_{1}) - G(\lambda_{0})
	=
	\int_{\lambda_{0}}^{\lambda_{1}} 
	\left<
	\frac{\ptl \Hamil(\lambdap)}{\ptl \lambdap}
	\right>
	\mathrm{d} \lambdap,
	\label{eq:TI_NpT_general}
\end{align}
where $\left<
\frac{\ptl \Hamil(\lambdap)}{\ptl \lambdap}
\right>$
denotes the equilibrium ensemble average of 
$\frac{\ptl \Hamil(\bm{\Gamma}, \lambdap)}{\ptl \lambdap}$
in the phase space of $\bm{\Gamma}$
which is substituted by the time average 
in the EMD systems in this study. 
%
%
%
By embedding $\lambda$ into the Hamilitonian 
in the right-hand side (RHS) so that $\Hamil(\bm{\Gamma}, \lambda)$ 
can analytically 
be differentiable by $\lambda$, the free energy difference 
in Eq.~\eqref{eq:TI_NpT_general} can be calculated by numerically integrating 
$\angb{\frac{\ptl \Hamil}{\ptl \lambdap}}$ obtained in each equilibrium system with a discrete $\lambdap$ value between $\lambda_{0}$ and $\lambda_{1}$. 
\par
Two implementations of the TI are used, 
the phantom wall (PW) method~\cite{Leroy2009, Leroy2010} and the dry surface (DS) method.~\cite{Leroy2015}
This provides a comparison of the two approaches as well as ensuring the TI is performed consistently.
Conceptually, the PW works like a pair of nets, pulled through the fluid each catching only one particle type to separate the two fluids. Meanwhile, the DS slowly changes the interaction of the two fluids, encouraging them to separate.   
\par
In the PW method, a wall is introduced which interacts only with the fluid particles through a short-range
repulsive potential function. This is applied to a quasi-1D 
system with a flat SL interface. 
The PW is set parallel to the interface, and the liquid 
is expelled by quasi-statically moving the PW starting 
from the solid side and moving to the liquid side under constant 
$NpT$ 
condition.
In this method, the PW position is linked with the 
coupling parameter $\lambda$ in the system Hamiltonian, 
and $\frac{\ptl \Hamil}{\ptl \lambdap}$ 
corresponds to the force exerted by the PW on the system, 
\ie the quasi-static work exerted on the system is 
calculated by the integral in Eq.~\eqref{eq:TI_NpT_general}.
This thermodynamic minimum work corresponds to the 
free energy difference between a system with a 
target SL interface at $\lambda = \lambda_{1}$ 
to a reference system with a solid surface exposed to vacuum 
and a PW-liquid interface achieved at 
$\lambda = \lambda_{0}$.~\cite{Surblys2014, Bistafa2021} 
\par
On the other hand, for the DS method, the parameter $\lambda$ is embedded into the SL interaction potential energy, and the free energy difference 
of the target system relative to the reference system with a ``dry" 
solid surface, in which the SL interaction 
is very weak or almost repulsive.~\cite{Surblys2018, Yamaguchi2019, Kusudo2019} In this method, $\angb{\frac{\ptl \Hamil}{\ptl \lambdap}}$ in Eq.~\eqref{eq:TI_NpT_general} corresponds to the total SL 
interaction energy of the system.
\par
Although the two methods give similar results, 
the DS method allows parameterization of the SL interfacial tension as a function of fluid interaction parameter with a lower computational cost than the PW method.~\cite{Surblys2018, Yamaguchi2019} 
This point is in contrast to the PW which needs the SL stripping 
process for each SL interaction parameter; however, the PW 
method is simple and therefore applicable to charged systems  including long-range interactions. In addition, 
the PW method gives a direct intuitive link with the thermodynamic 
minimum work.
\par
In this study, we used the DS and PW methods both extended 
for the liquid-liquid interface.
%
%
%
In both methods, we evaluated the thermodynamic 
minimum work needed to change from a target system to 
a reference system with $\alpha$ and $\beta$ completely 
isolated without mixing where the contribution to the free energy difference vanished. 
Note that the implementations of the reference systems 
were different for the DS and PW methods 
as in Figs.~\ref{fig:system}~(b-i) and (b-ii);
however, they are 
assumed equivalent from a 
thermodynamic point of view.
\par
The details of the two methods are described in Appendix~\ref{appsec:TI}. 
The basic point is that the thermodynamic equilibrium state of the present $NpT$ constant system is determined by giving two variables of miscibility $\eta$ and the phantom wall position $\xpw$ which corresponded to the positions of the symmetric semipermeable PWs at $\pm \xpw$, and we change only one as the coupling parameter for the TI with keeping the other unchanged. Note that the average piston positions $\left<\xpcl\right>$ and $\left<\xpcr\right>$ are dependent variables determined by $(\eta, \xpw)$ through the control pressure $p$.
In the DS method, we set the miscibility parameter $\eta$ 
in Eq.~\eqref{eq:lj_potential_eta} as the coupling 
parameter for the TI with keeping $\xpw=\xpw_{1}$ unchanged, 
and reproduced the reference system by setting $\eta \rightarrow 0$ keeping $\eta$ positive as in the original DS method.~\citep{Leroy2015}
In the present case, we set the system at $\eta=\eta_{0}\ (=0.01)$ 
as the reference system with completely isolated 
liquids as shown in Fig.~\ref{fig:system}~(b-i). The value of $\xpw_{1}$ was set large enough that the PWs were located far away from the liquid (behind the pressure control pistons) and did not interact with the fluid particles, \ie the PWs had no contribution to the system Hamiltonian for the present DS procedure.
Then, based on Eq.~\eqref{eq:TI_NpT_general}, the 
free energy difference between the target system 
at $(\eta,\xpw)=(\eta,\xpw_{1})$ 
and the reference system at
$(\eta_{0},\xpw_{1})$ 
is calculated by
\begin{align}
	G(\eta,\xpw_{1})- G(\eta_{0},\xpw_{1})
	=
	\int_{\eta_{0}}^{\eta} 
	\angb{
		\frac{\partial \Hamil(\eta',\xpw_{1})}{\partial \eta'} 
	} \mrd \eta',
	\label{eq:def_DeltaG_DS}
\end{align}
where the integrand $\left<\frac{\partial \Hamil(\eta',\xpw_{1})}{\partial \eta'}\right>$ in the RHS can be 
easily obtained in the MD system as in the original DS procedure.
According to the second law of thermodynamics, this corresponds
to the sum of the minimum work  
needed for the change from the target system 
to the reference system 
and the quasi-static work exerted on the pressure control pistons as the environment
under constant $NpT$.
By separating the work on the PCs per unit area 
given by 
$\Wpc\left[(\eta,\xpw_{1})\rightarrow(\eta_{0},\xpw_{1})\right]$ (see Eq.~\eqref{app_eq:def_Wpc_DS} for the definition), 
we define $\WisoDS(\eta)$ as the ``work for 
isolation" calculated by the DS method as
\begin{align}
	\WisoDS(\eta)
	\equiv
	-\frac{G(\eta,\xpw_{1})- G(\eta_{0},\xpw_{1})}{A} 
	- 
	\Wpc\left[(\eta,\xpw_{1})
	\rightarrow
	(\eta_{0},\xpw_{1})\right].
	\label{eq:def_Wiso_DS}
\end{align}
\par
In this study, we prepared multiple equilibrium systems 
with different miscibility parameter $\eta\in[0.01,1]$, 
and calculated the time average 
of the average LL potential energy over 20~ns 
for each equilibrium system to numerically integrate
the RHS of Eq.~\eqref{eq:def_DeltaG_DS} 
(see Eq.~\eqref{app_eq:w_iso_DS} in Appendix~\ref{appsec:TI}). 
With this procedure, dependence of $\WisoDS(\eta)$ on $\eta$ was clearly observed through the numerical integration with respect to $\eta$.
\par
On the other hand, in the PW method, we used 
$\xpw$ as the coupling parameter with keeping $\eta$ unchanged, and 
reproduced the reference system by pushing the PWs 
as in Fig.~\ref{fig:system}~(b-ii), 
\ie decreasing $\xpw$ from $\xpw_{1}$ down to 
$\xpw_{0}$ so that the two liquids were completely 
isolated. 
%
In this case, the difference of $G(\eta, \xpw)$ is written by
\begin{align}
	G(\eta,\xpw_{1})- G(\eta,\xpw_{0})
	=
	\int_{\xpw_{0}}^{\xpw_{1}} 
	\angb{
		\frac{\partial \Hamil(\eta,\xpwp)}{\partial \xpwp} 
	} \mrd \xpwp.
	\label{eq:def_DeltaG_PW}
\end{align}
It is obvious from the PW-fluid potential 
functions in Eqs.~\eqref{eq:potential_pwa}, 
\eqref{eq:potential_pwb} and \eqref{eq:xpwb=-xpwa} 
that the Hamiltonian derivative is reduced to 
the forces on the PWs as
\begin{align}
	\left<
	\frac{\partial \Hamil(\eta, \xpw)
	}{
		\partial \xpw}
	\right>
	=
	-\angb{
		F_{\mathrm{fluid-pw}\alpha}(\eta, \xpw)
	}
	+\angb{
		F_{\mathrm{fluid-pw}\beta}(\eta, \xpw)
	},
	\label{eq:ptl_H_ptl_xpw}
\end{align}
where 
$\angb{
	F_{\mathrm{fluid-pw}\alpha}(\eta, \xpw)
}$
and 
$\angb{
	F_{\mathrm{fluid-pw}\beta}(\eta, \xpw)
}$
denote the forces on the phantom-walls 
$\alpha$ and $\beta$ from the corresponding 
liquids, respectively. Note that 
\begin{equation}
	\angb{F_{\mathrm{fluid-pw}\alpha}}
	\approx
	-
	\angb{F_{\mathrm{fluid-pw}\beta}}
	\geq 0
\end{equation}
for the forces because of the repulsive setting.
Hence, the RHS of 
Eq.~\eqref{eq:def_DeltaG_PW} corresponds to 
the work exerted by the two PWs (see 
Appendix~\ref{appsec:TI} for details).
Similar to Eq.~\eqref{eq:def_Wiso_DS}, 
we define $\WisoPW(\eta)$ as the work for 
isolation calculated by the PW method as
\begin{align}
	\WisoPW(\eta)
	\equiv
	-\frac{G(\eta,\xpw_{1})- G(\eta,\xpw_{0})}{A} 
	- 
	\Wpc\left[(\eta,\xpw_{1})
	\rightarrow
	(\eta,\xpw_{0})\right]
	\label{eq:w_iso_PW}.
\end{align}
Note that the PW position $\xpw$ was varied while 
keeping $\eta$ constant.
In practice, we calculated $\WisoPW(\eta)$ 
for several discrete $\eta$ values. This was
in contrast to the calculation of $\WisoDS(\eta)$, 
which was calculated through the numerical integration 
with respect to $\eta$. 
\par
We compared $\WisoPW(\eta)$, 
$\WisoDS(\eta)$ and the interfacial tension $\gab(\eta)$ for 
various $\eta$ values. We validated that the two TIs were carried out quasi-statically with tracing equilibrium points for both reversible paths through the comparison between $\WisoPW(\eta)$ and $\WisoDS(\eta)$. This will be described in 
\ref{subsec:comp_tens_works} with Fig.~\ref{fig:diff}.
\section{Results and discussion}
\label{sec:resdis}
\subsection{Stress distribution and resulting interfacial tension}
\label{subsec:stress_dis & Delta_F}
\begin{figure}[t]
	\begin{center}
		\includegraphics[scale=0.5]{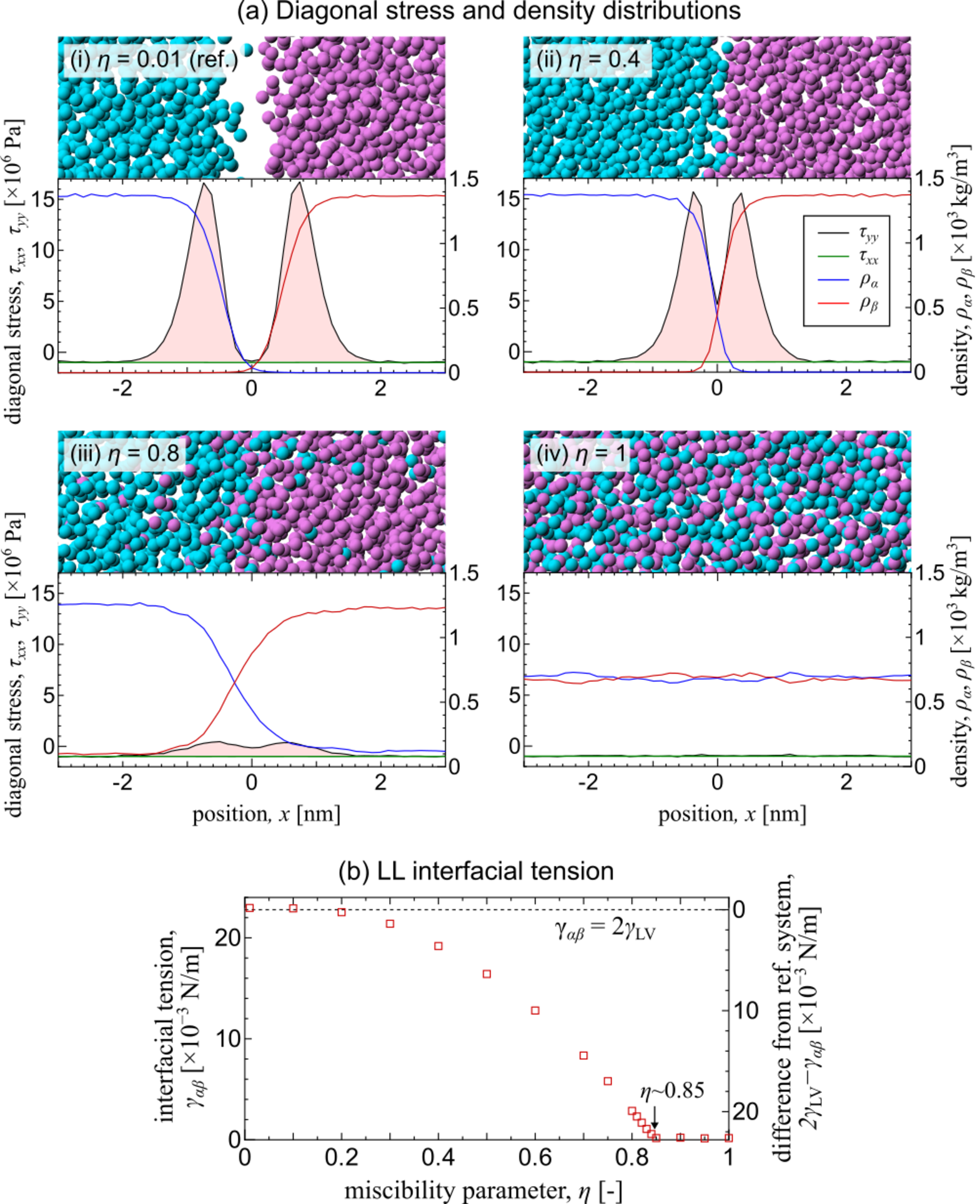}
	\end{center}
	\caption{
		\label{fig:stress_dis}		
		(a) Distributions of the diagonal stress components 
		$\tau_{yy}$ (black) and $\tau_{xx}$ (green) 
		calculated by the VA, 
		and the densities $\rho_{\alpha}$ (blue) and 
		$\rho_{\beta}$ (red) of $\alpha$ and $\beta$ 
		components for the systems at $\eta =0.01$, 
		0.4, 0.8 and 1.
		Enlarged snapshots of the systems around the 
		interface are also shown. (b) Interfacial tension $\gab$
		calculated from the stress distribution by 
		Eq.~\eqref{eq:ex_Bakker's eq}
		as the mechanical route (MR). The value of $2\glv$ 
		obtained in an independent system is also displayed.
	}
\end{figure}
Figure~\ref{fig:stress_dis}~(a) shows the distributions of  
the diagonal stress components $\tau_{yy}$ and $\tau_{xx}$ calculated by the VA and the density distributions of 
each component for the systems at $\eta =0.01$, 
$0.4$, $0.8$ and $1$. Enlarged snapshots of the systems around the 
interface are shown above the distributions.
As observed in the density distributions and snapshots, 
two liquids completely isolated at $\eta=0.01$ were 
gradually mixed with the increase of $\eta$, and one 
component dissolved into the other at $\eta=0.8$ 
where for instance, $\rho_\alpha$ was non-zero even in the right side of away from the interface. 
The two liquids were completely mixed at $\eta=1$ 
and the density was homogeneous without forming an interface 
because the two particles were identical.
Regarding the stress distributions, 
by using the stress definition described in 
Sec.~\ref{subsec:mechanical route}, the uniformity of $\tau_{xx}$ 
in Eq.~\eqref{eq:tau=const}, which 
was consistent with mechanical equilibrium condition 
in the direction normal to the interface, was satisfied 
for all $\eta$ values even at the interfaces with 
a steep spatial change in the densities. 
This indicates that the present 
definition including all component's kinetic contributions  
$\bm{\tau}^\mathrm{kin}$ and all 
interactions between particles 
of both the same and different components to the 
interaction term $\bm{\tau}^\mathrm{int}$ respects the condition of mechanical equilibrium.
In addition, the constant value $\tau_{xx}$ was equal
to $\tau_{yy}$ in the bulk regions away from the 
interface, \ie the stress isotropy was satisfied there. The isotropic value $\tau_{xx} (=\tau_{yy})$ in bulk is negative because this is equal to the bulk pressure with its sign reverted.
Based on these results, we obtained the LL interfacial 
tension $\gab(\eta)$ by Eq.~\eqref{eq:ex_Bakker's eq} 
as a function of $\eta$ using the present definition. 
This corresponds to the integral for the regions 
filled with light red in Fig.~\ref{fig:stress_dis}~(a).
\par
Two distinct regions with $\tau_{yy} > \tau_{xx}$ 
separated at the boundary of two liquids existed 
for $\eta=0.01 (=\eta_{0})$ [Fig.~\ref{fig:stress_dis}~(a-i)], 
where the two liquids were isolated 
and the sum of the densities $\rho_\alpha + \rho_\beta$
were almost zero at the boundary.
The interface is equivalent to the system with two 
interfaces between the liquid (L) 
and vapor (V) each with a surface tension of $\glv$, 
\ie 
\begin{equation}
	\gab(\eta_{0})
	\approx
	2\glv.
	\label{eq:gab@eta=0}
\end{equation}
This can also be intuitively understood from the distribution 
of $\tau_{yy}$ with two peaks displayed in the figures Fig.~\ref{fig:stress_dis}~(a-i).
Indeed, we independently performed an EMD simulation of a quasi-1D 
single component LV system at coexistence with two flat LV interfaces 
at the same temperature and calculated the value of $\glv$,~\cite{Yamaguchi2019} 
and confirmed that the resulting $2\glv$ 
was consistent with the 
value of $\gab(\eta_{0})$ at $22.96\pm0.03\times 10^{-3}$~N/m 
in the present study as indicated in Eq.~\eqref{eq:gab@eta=0}.
Note that the transverse stress $\tau_{yy}$ reached very large positive values $\sim +10^{7}$~Pa, as estimated by \citet{Rowlinson1982}, in comparison to the bulk value around $-10^{6}$~Pa.
\par
The two isolated peaks of $\tau_{yy}$ were merged with the 
increase of $\eta$ and the maximum values of $\tau_{yy}$ 
became smaller, thus, the resulting integral was also reduced
[Fig.~\ref{fig:stress_dis}~(a-ii,iii)]. 
At $\eta=1$, the two stress components $\tau_{yy}$ and
$\tau_{xx}$ were equal and the corresponding integral 
in Eq.~\eqref{eq:ex_Bakker's eq} was zero
[Fig.~\ref{fig:stress_dis}~(a-iv)]. 
This physically means that no interface exists and 
the mechanical interfacial tension satisfies
\begin{equation}
	\gab(\eta=1) = 0.
	\label{eq:gab@eta=1}
\end{equation}
Figure~\ref{fig:stress_dis}~(b) shows $\gab(\eta)$ 
calculated by Eq.~\eqref{eq:ex_Bakker's eq} for 
each value of $\eta$. As expected, $\gab(\eta)$ 
monotonically decreased with the increase of 
$\eta$, and it reached zero at around $\eta=0.85$.
Above this value, the two liquids were mixed 
and no LL interface were formed as shown in 
Fig.~\ref{fig:system}~(b-i).
We used $-[\gab(\eta)-2\glv]$ as the difference 
from $\gab(\eta_{0})$ indicated in the right vertical
axis for the comparison with the works for isolation
below.
%
%
%
\subsection{Comparison of interfacial tension and works for isolation}
\label{subsec:comp_tens_works}
We compared the mechanical interfacial tension $\gab(\eta)$ and the 
works for isolation 
$\WisoDS(\eta)$ and $\WisoPW(\eta)$
obtained by the DS and extended-PW methods, respectively
for various $\eta$. 
Prior to that, we set the standard basis for the comparison 
through the difference of $\gab(\eta)$ and $\Wiso(\eta)$ 
considering two representative cases of $\eta=\eta_{0}$ 
and $\eta=1$ 
where the values could be evaluated from the physical 
meanings. As shown in Fig.~\ref{fig:stress_dis},
$\gab(\eta_{0})=2\glv$ and $\gab(1)=0$ in 
Eqs.~\eqref{eq:gab@eta=0} and \eqref{eq:gab@eta=1}, 
respectively hold for $\gab(\eta)$.
Meanwhile for the works for isolation, it is 
obvious that 
\begin{equation}
	\WisoDS(\eta_{0})=0
	,\quad\mathrm{and}\quad  
	\WisoPW(\eta_{0})=0,
	\label{eq:Wsys@eta=0}
\end{equation}
where the second equality is because no work is given 
to the system for a further separation of 
the two isolated liquids by the PWs.
%
On the other hand, for the case of $\eta=1$ where 
$\alpha$ and $\beta$ are identical, 
a certain positive work $\WisoPW(\eta=1)$
is needed to isolate $\alpha$ and $\beta$ from 
this mixed state 
although the correspondence with the mechanical 
interfacial tension $\gab(1)$ is not clear.
%
Considering these points,
we compared $-[\gab(\eta)-2\glv]$ using the 
values obtained by the mechanical route 
with $\Wiso(\eta)$ obtained by the two thermodynamic routes.
%
Note that the value $2\glv$ corresponds to the work per 
unit area required to divide a single component liquid 
into two isolated parts with a depletion region 
equivalent to the gas phase between them as in the 
reference system.
Also note that the unit 
N/m for the mechanical route is 
equivalent to the unit J/m$^{2}$ for the thermodynamic 
route.
\par
\begin{figure}[t]
	\begin{center}
		\includegraphics[scale=0.80]{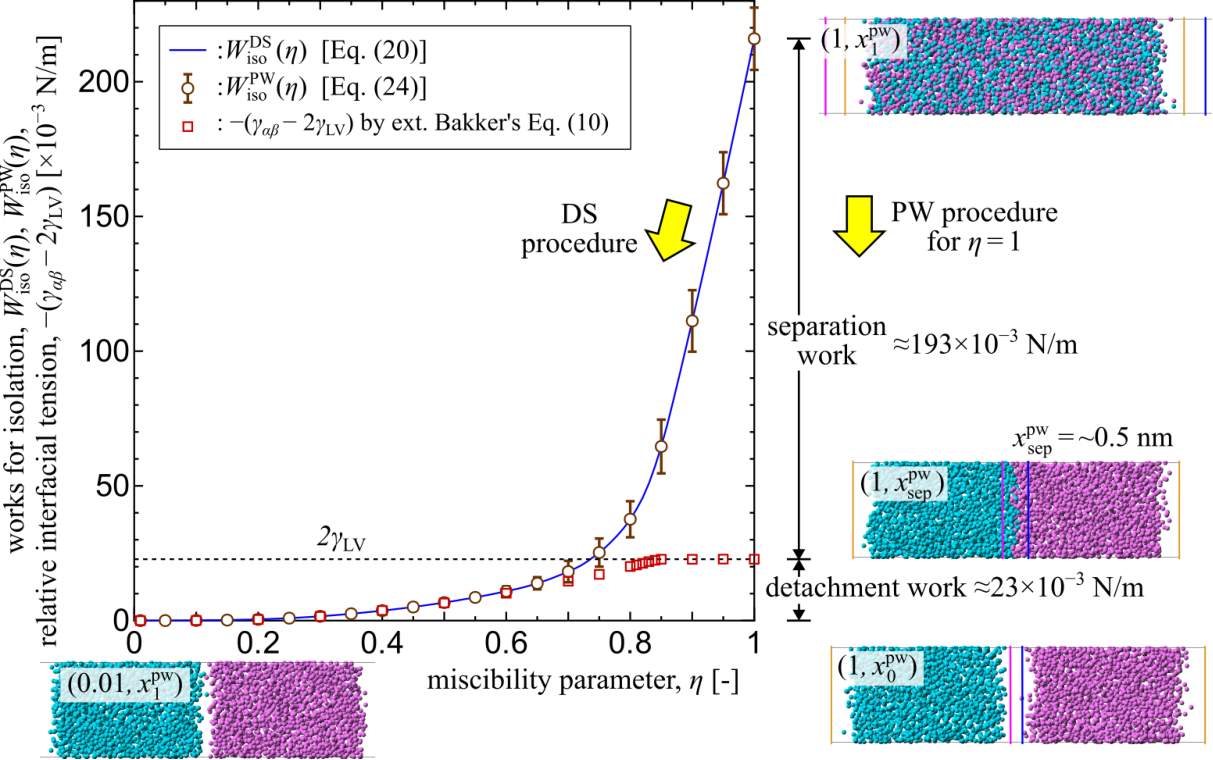}
	\end{center}
	\caption{
		\label{fig:diff}		
		Comparison between the relative interfacial tension 
		$-(\gab - 2\glv)$ obtained with the mechanical route and work for isolation $\Wiso$
		for various $\eta$. Error bars for $\WisoDS$ are not shown for better visualization here (see Fig.~\ref{fig:result_DS}).}
	Side snapshots in Fig.~\ref{fig:system}~(b) 
	with the corresponding $(\eta, \xpw)$ value
	are appended for some systems.
\end{figure}
Figure~\ref{fig:diff} displays the comparison among 
relative interfacial tension 
$-[\gab(\eta)-2\glv]$ and the works for isolation 
$\WisoDS(\eta)$ and $\WisoPW(\eta)$ 
as a function of 
the miscibility parameter $\eta$. 
Note that the error bars for $\WisoDS$ are not shown in this figure for better visualization; however, they are relatively small as shown in Fig.~\ref{fig:result_DS} in Appendix~\ref{appsec:DS_result}.
We start from the comparison between 
$\WisoDS(\eta)$ and $\WisoPW(\eta)$. As described in Subsec.~\ref{subsec:thermodynamic route}, 
$\WisoDS(\eta)$ was obtained as a quasi-smooth function 
by the extended-DS method.
On the other hand, $\WisoPW(\eta)$ was obtained 
for discrete $\eta$ values.
Nevertheless, it was shown that the two values 
$\WisoDS(\eta)$ and $\WisoPW(\eta)$ gave the 
same result although the thermodynamic 
integration paths were toward equivalent but 
different reference systems 
at $(\eta, \xpw)=(\eta_{0}, \xpw_{1})$ and $(\eta, \xpw_{0})$, respectively. 
This match also indicates that the works of isolation $\Wiso$ were correctly obtained by the two TIs by quasi-statically tracing the equilibrium thermodynamic points along the reversible TI paths.
Note that the error bars were larger for 
the PW method because the force on the PWs in 
Eq.~\eqref{eq:ptl_H_ptl_xpw} was used 
which was subject to larger thermal fluctuations
(see Fig.~\ref{fig:PW_eta=1}). 
\par
Regarding the comparison between $-(\gab(\eta) - 2\glv)$ 
obtained by the mechanical route and $\Wiso(\eta)$ 
by the thermodynamic routes, they matched well for small 
$\eta$ values; however, with the increase of $\eta$ from about 
$\eta=0.7$, the difference became large and it steeply 
increased above $\eta=0.8$. 
For instance, the difference for the completely mixing case at $\eta=1$
was about $193\times 10^{-3}$~N/m, which was much 
larger than $-(\gab - 2\glv)$ of about $23\times 10^{-3}$~N/m. 
Briefly, the difference was because an extra work 
is needed to separate the components in the bulk liquid regions in addition 
to completely detach the two fluids into two parts,
as schematically illustrated in the right of 
Fig.~\ref{fig:diff}. This point will be further examined 
in the next subsection~\ref{subsec:decomp_wiso}. 
%
%
\subsection{Decomposition of the work for isolation for the completely miscible case
}
\label{subsec:decomp_wiso}
\begin{figure}[t]
	\begin{center}
		\includegraphics[width=0.6\linewidth]{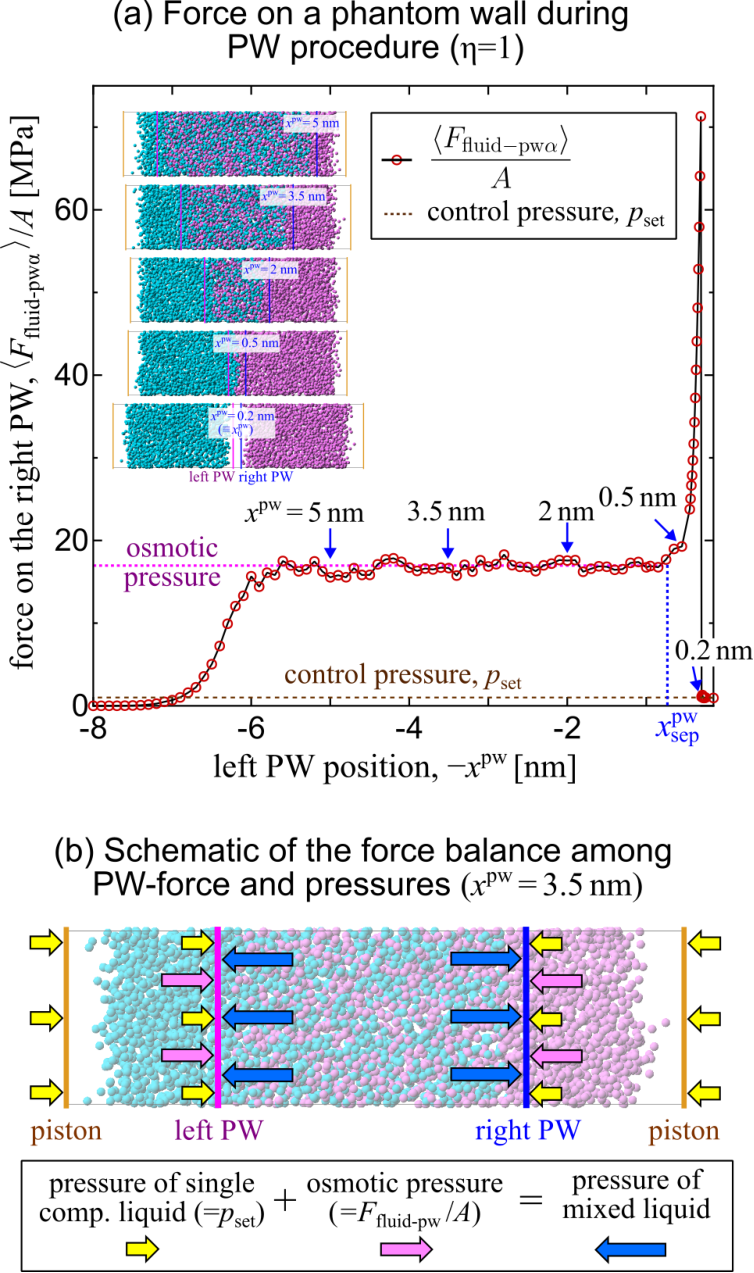}
	\end{center}
	\caption{
		\label{fig:PW_eta=1}		
		(a) Force on the right phantom wall (PW) 
		per unit area upon the calculation of the 
		work for isolation $\WisoPW(\eta)$ by 
		the extended-PW method in the completely 
		miscible case ($\eta=1$). 
		The inset corresponds to 
		Fig.~\ref{fig:system}~(a-ii).
		(b) Schematic of the force balance among 
		the force on the left and right PWs,  
		the pressure of the two single component 
		liquids, and that of the mixed liquid in the center between the two PWs.
	}
\end{figure}
To examine the difference between 
the mechanical interfacial tension and the thermodynamic 
work for isolation observed in Fig.~\ref{fig:diff}, 
we looked into the intermediate process of the 
PW method in the completely miscible case 
($\eta=1$), focusing on the force 
on the PWs.
Figure~\ref{fig:PW_eta=1}~(a) shows the force 
on the right phantom wall (PW) per unit area 
$\frac{\angb{F_{\mathrm{fluid-pw}\alpha}}}{A}$ 
upon the calculation of the work for isolation 
$\WisoPW(\eta)$ by the PW method at 
$\eta=1$, 
where the inset corresponds to Fig.~\ref{fig:system}~(a-ii).
Note that $\alpha$-$\beta$ interface were not 
formed in the $\alpha$-$\beta$ 
mixture because the two liquids were identical 
in this system at $\eta=1$.
As the PW entered into the fluid with the 
increase of 
$\xpw$, almost constant force was exerted on the 
PW as observed at $\xpw=5$, $3.5$ and 2~nm, where 
the liquid separation by the semipermeable PWs 
proceeded with reducing the volume of the mixed liquid.
The force $\frac{\angb{F_{\mathrm{fluid-pw}\alpha}}}{A}$ 
was remarkably larger than the control pressure at 
these states.
This force corresponds to the osmotic pressure,~\cite{Luo2010,Bley2017} as 
will be discussed in Subsec.~\ref{subsec:osmop}.
The force steeply 
rose up just before the fluid was completely isolated 
at $\xpw=0.5$~nm, and it decreased down to the control 
pressure at $\xpw=0.2$~nm. 
No work was needed to further separate the 
liquids where the work done by the PWs to the system 
and that on the PCs balanced.
\par
As indicated with the arrows and snapshots in Fig.~\ref{fig:diff}, 
the thermodynamic integration was started from 
$(\eta,\xpw)=(1,\xpw_{1})$ to equivalent different 
systems, \ie to $(\eta_{0},\xpw_{1})$ for the DS and 
to $(\eta,\xpw_{0})$ for the PW methods. 
Indeed along the path of the PW method, the liquids 
were gradually separated until the intermediate 
state having a LL interface without sandwiching 
vapor as the snapshot at $\xpw=0.5$~nm in the inset of Fig.~\ref{fig:PW_eta=1}.
From this intermediate state, the two liquids were 
completely detached by the PWs to achieve the 
reference system having two isolated LV interfaces.
Now, for the case at $\eta=1$ where $\alpha$ 
and $\beta$ were identical, we assume $\gab$ 
to be zero at the intermediate state even 
though two liquids are separated by the 
semipermeable PWs. 
Then, 
let $\xpw_\mathrm{sep}(\eta)$ be the PW position of the intermediate state, 
the minimum work 
$\Delta W\left[(\eta,\xpw_\mathrm{sep}(\eta))
\rightarrow
(\eta,\xpw_{0})\right]
|_{\eta=1}$
needed for the change from this intermediate 
system at $(\eta,\xpw)=(1,\xpw_\mathrm{sep}(1))$ 
to the reference system at $(1,\xpw_{0})$ is equal 
to $2\glv$, \ie
\begin{equation}
	\Delta W\left[(\eta,\xpw_\mathrm{sep}(\eta))
	\rightarrow
	(\eta,\xpw_{0})\right]
	|_{\eta=1} 
	= 
	-[\gab(\eta=1)-2\glv]
	= 2\glv,
\end{equation}
Note that $\xpw_\mathrm{sep}(\eta)|_{\eta=1}$ was about 0.5~nm as indicated in Fig.~\ref{fig:PW_eta=1}, but in general, $\xpw_\mathrm{sep}(\eta)$ is given as a function of $\eta$ because the position slightly depends on the mixing feature governed by $\eta$.
We define this as the `detachment work' $\Wdet(\eta)$ as
\begin{equation}
	\Wdet(\eta) 
	\equiv 
	\Delta W\left[(\eta,\xpw_\mathrm{sep}(\eta))
	\rightarrow
	(\eta,\xpw_\mathrm{0})\right],
	\label{eq:def_Wdet}
\end{equation}
which satisfies
\begin{equation}
	\Wdet(1) 
	=
	2\glv
	\label{eq:Wdet@eta=1}
\end{equation}
for a specific case at $\eta = 1$.
In addition, we introduce the minimum work from 
the target system $(\eta,\xpw_{1})$ to the intermediate 
state at $(\eta,\xpw_\mathrm{sep}(\eta))$ 
to separate the liquids defined by
\begin{equation}
	\Wsep(\eta)
	\equiv
	\Delta W\left[(\eta,\xpw_\mathrm{1})
	\rightarrow
	(\eta,\xpw_\mathrm{sep}(\eta))\right]
	=
	\Wiso(\eta) - \Wdet(\eta),
	\label{eq:def_Wsep}
\end{equation}
which we call the `separation work.'
\par
The value of $\Wsep(1)$ is estimated  
for a specific case at $\eta=1$ from a viewpoint of configurational entropy here. 
Indeed, $\eta=1$ corresponds to an ideal mixture for which the mixing free energy is purely of entropic origin. More in detail,
since $\alpha$ and $\beta$ were identical under the constant temperature $T$ and pressure $p$ in this
case, the two states should have the same liquid structure with the same internal energy $U$ and volume $V$ even though fluids $\alpha$ and $\beta$ were separated on the left and right sides, respectively.
In other words, the difference between the two states is:
particles of $\alpha$ and $\beta$ can move in the whole space 
between the two PCs in the mix-state whereas each kind of particles 
can move only half of the space.
Let $G$ and $S$ be the Gibbs free energy and entropy, respectively 
and let `mix' and `sep' be the subscripts for completely mixed 
target and intermediate states, respectively.
Then, it follows for $\Wsep(1)$ that
\begin{align}
	\nonumber
	\Wsep(1) 
	&= 
	\frac{1}{A}\left(
	G_\mathrm{sep} - G_\mathrm{mix}
	\right)|_{\eta=1}
	\\
	\nonumber
	&=
	\frac{1}{A}\left[
	(U_\mathrm{sep}+pV_\mathrm{sep}-TS_\mathrm{sep}) 
	-
	(U_\mathrm{mix}+pV_\mathrm{mix}-TS_\mathrm{mix}) 
	\right]|_{\eta=1}
	\\
	&=
	\frac{T}{A}(S_\mathrm{mix}-S_\mathrm{sep})|_{\eta=1},
	\label{eq:Wsep_thermodyn}
\end{align}
because 
$U_\mathrm{mix}=U_\mathrm{sep}$ 
and 
$V_\mathrm{mix}=V_\mathrm{sep}$
are assumed. 
The entropy difference can be estimated by the possible 
volume available for liquid $\alpha$ and $\beta$ composed of
$N_\alpha$ and $N_\beta$ particles, respectively, 
as in a thought experiment of ideal gas separation by
using semipermeable membrane in ordinary thermodynamics 
and statistical mechanics textbooks~\cite{Greiner1995} as follows:
\begin{align}
	\nonumber
	(S_\mathrm{mix}-S_\mathrm{sep})|_{\eta=1}
	&\equiv
	\left(S_\mathrm{mix}^\alpha-S_\mathrm{sep}^\alpha\right)|_{\eta=1}
	+
	\left(S_\mathrm{mix}^\beta-S_\mathrm{sep}^\beta\right)|_{\eta=1}
	\\
	\nonumber
	&=
	\left.
	N_\alpha \kB 
	\ln \frac{V_\mathrm{mix}}{V^\alpha_\mathrm{sep}}
	\right|_{\eta=1}
	+
	\left.
	N_\beta \kB 
	\ln \frac{V_\mathrm{mix}}{V^\beta_\mathrm{sep}}
	\right|_{\eta=1}
	\\
	&=
	2N_\alpha \kB
	\left.
	\ln{\frac{V_\mathrm{mix}}{V^\alpha_\mathrm{sep}}}
	\right|_{\eta=1}
	\label{eq:S_eval}
\end{align}
where $k_\mathrm{B}$ is the Boltzmann constant, and $V_\mathrm{mix}$, $V^\alpha_\mathrm{sep}$ and $V^\beta_\mathrm{sep}$ are the 
volumes of the liquid in the mixed state and those for $\alpha$ 
and $\beta$ parts in the intermediate state. 
Note that $N_\alpha=N_\beta$ and $V_\alpha = V_\beta$ were 
used for the final equality. 
The resulting $\Wsep(1)$ obtained by Eq.~\eqref{eq:def_Wsep} 
using the DS result of $\WisoDS(1)$ and 
$\Wdet(1) = 2 \glv$ was 
\begin{equation}
	\Wsep(1)
	=
	\Wiso^\mathrm{DS}(1) - 2\glv, 
	=
	(193.0 \pm 1.0) \times 10^{-3}\mathrm{\ J/m^{2}}.
	\label{eq:Wsep_eta=1}
\end{equation}
On the other hand, the entropy difference 
estimated by Eqs.~\eqref{eq:Wsep_thermodyn}
and \eqref{eq:S_eval} was
\begin{equation}
	\frac{2N_\alpha k_\mathrm{B} T}{A} 
	\left.
	\ln{\frac{V_\mathrm{mix}}{V_\mathrm{sep}^\alpha}}
	\right|_{\eta=1}
	=(185.7\pm 9.6) \times 10^{-3}\mathrm{\ J/m^{2}},
\end{equation}
where the PW position and average positions of 
the PCs for the system at 
$(\eta,\xpw)=(1, 0.5\mathrm{\ nm})$ 
were used to roughly estimate the volume $V_\mathrm{sep}^\alpha$.
Indeed, the two agreed well, and this indicated that 
the work for isolation included the interfacial tension 
and the mixing free energy. 
Note that the method can also be used for non-ideal mixtures, \ie for the case of $\eta \neq 1$, to extract both the interfacial free energy change and the free energy of mixing, which will then contain both an enthalpic and an entropic contributions.


%
\subsection{Osmotic pressure
}
\label{subsec:osmop}
We discuss the meaning of the constant force per 
unit area $\frac{\angb{F_{\mathrm{fluid-pw}\alpha}}}{A}$,
which was larger than the control pressure,
observed during the liquid separation in Fig.~\ref{fig:PW_eta=1}~(a).
Figure~\ref{fig:PW_eta=1}~(b) illustrates the schematic
of the force balance at $\xpw=3.5$~nm.
The pressure was controlled at $p_\mathrm{set}$
by the pressure control pistons on both ends of the 
system. This indicated that the two single component 
liquids in the left and right both 
between a piston and a phantom wall (PW) had a pressure 
of $p_\mathrm{set}$. On the other hand, the mixed 
liquid between the PWs in the center was subject to the pressure 
of the single component liquids as well as the PWs.
Hence, the constant force per unit area shown with dotted purple 
line in Fig.~\ref{fig:PW_eta=1}~(a) corresponded to
the osmotic pressure, as discussed in previous work.~\cite{Luo2010,Bley2017}

\section{Conclusion}
We performed molecular dynamics simulations of a liquid-liquid interface between two different Lennard-Jones liquids with various miscibility and evaluated the interfacial tension using both mechanical and thermodynamic routes.
In the case of the mechanical route, the vertical stress normal to the interface was observed to be constant over the entire region provided all kinetic and interaction contributions were included in the stress. 
From these stress distributions, we calculated the liquid-liquid interfacial tension obtained by using Bakker's equation for various miscibility and compared with the free energy obtained by the thermodynamic routes, where the extended dry-surface and phantom-wall schemes were used to quasi-statically isolate the two liquids under a constant pressure and temperature condition.
When the two components were immiscible, the mechanical and thermodynamic results were in good agreement whereas when they were miscible, the values were significantly different. 
This difference was attributed to the additional free energy required 
to separate the binary liquid into two components, \ie the free energy of mixing. In the phantom wall setup, it was possible to disentangle the free energy of mixing, which corresponded to the work of the osmotic pressure acting on the phantom wall prior to the complete detachment of the two components, and the change in interfacial free energy occuring upon detachment. 
In the ideal mixture case, we showed that the free energy of mixing corresponded to the entropy difference between mixed state and separated state. For non-ideal mixtures, the PW method provides the full free energy of mixing -- including an enthalpic and an entropic contributions, together with the osmotic pressure of the mixtures.
\begin{acknowledgments}
We thank Haruki Oga for fruitful discussion as a former member of 
Y.Y.'s group at Osaka University.
H.K., T.O., and Y.Y. were supported by JSPS KAKENHI grant (Nos. JP23KJ0090, JP23H0134, and 22H01400), Japan, respectively. 
Y.Y. was also supported by JST CREST grant (No. JPMJCR18I1), Japan.
We discussed this research during the NEMD Conference held on 13th and 14th, June 2024, partly supported by JSPS Bilateral Joint Research Seminars (Japan-UK, No. 220249903).
\end{acknowledgments}
%
\vspace{5mm} \par \noindent
\textbf{DATA AVAILABILITY}
\par
The data that support the findings of this study are available from the corresponding author upon reasonable request.
\vspace{5mm} \par \noindent
\textbf{AUTHOR DECLARATIONS}
\newline
\textbf{Conflict of Interest}
\par
The authors have no conflicts of interest to disclose.
\appendix
\section{Thermodynamic integration by the extended-DS and PW methods}
\label{appsec:TI}
%
The basic point is that the thermodynamic equilibrium state of the present $NpT$ constant system is determined by giving two variables of miscibility $\eta$ and the phantom wall position $\xpw$ which corresponded to the positions of the symmetric semipermeable PWs at $\pm \xpw$, and we change only one as the coupling parameter for the TI with keeping the other unchanged.
Note that the average piston positions $\left<\xpcl\right>$ and $\left<\xpcr\right>$ are dependent variables determined by $(\eta, \xpw)$ through the control pressure $p$.
In the DS method, we set the miscibility parameter $\eta$ 
in Eq.~\eqref{eq:lj_potential_eta} as the coupling 
parameter for the TI, 
and reproduced the reference system by setting $\eta \rightarrow 0$ 
with keeping $\eta$ positive similar to the DS method.
In the present case, we set the system at $\eta=\eta_{0}\ (=0.01)$ 
as the reference system with completely isolated 
liquids as shown in Fig.~\ref{fig:system}~(b-i).
Then, based on Eq.~\eqref{eq:TI_NpT_general}, the 
free energy difference between the target system 
at $(\eta,\xpw)=(\eta,\xpw_{1})$ 
and the reference system at
$(\eta_{0},\xpw_{1})$ 
is calculated by
\begin{align}
	G(\eta,\xpw_{1})- G(\eta_{0},\xpw_{1})
	=
	\int_{\eta_{0}}^{\eta} 
	\angb{
		\frac{\partial \Hamil(\eta',\xpw_{1})}{\partial \eta'} 
	} \mrd \eta'.
	\label{app_eq:def_DeltaG_DS}
\end{align}
According to the second law of thermodynamics, this corresponds
to the sum of the minimum work exerted on the PCs and that 
needed for the change from the target system 
to the reference system under constant $NpT$.
%
Let the two works both per unit area be defined by
$\Wpc\left[(\eta,\xpw_{1})
\rightarrow
(\eta_{0},\xpw_{1})\right]$ 
and 
$\Delta W\left[(\eta,\xpw_{1})
\rightarrow
(\eta_{0},\xpw_{1})\right]$, respectively, 
it follows for the former that
\begin{align}
	\nonumber
	\Wpc\left[(\eta,\xpw_{1})
	\rightarrow
	(\eta_{0},\xpw_{1})\right]
	&= 
	\pset\left[
	\angb{\xpcr(\eta_{0},\xpw_{1})}
	-
	\angb{\xpcr(\eta,\xpw_{1})}
	\right]
	\\
	&+
	\pset\left[
	\angb{\xpcl(\eta,\xpw_{1})}
	-
	\angb{\xpcl(\eta_{0},\xpw_{1})}
	\right].
	\label{app_eq:def_Wpc_DS}
\end{align}
Hence, the latter can be obtained by
%
\begin{align}
	\Delta W\left[(\eta,\xpw_{1})
	\rightarrow
	(\eta_{0},\xpw_{1})\right]
	= 
	-\frac{G(\eta,\xpw_{1})- G(\eta_{0},\xpw_{1})}{A} 
	- 
	\Wpc\left[(\eta,\xpw_{1})
	\rightarrow
	(\eta_{0},\xpw_{1})\right]
	\label{app_eq:def_Wsys_DS}
\end{align}
We define the ``work for isolation" by the DS 
denoted by $\WisoDS(\eta)$ as this difference 
in this study, \ie
\begin{equation}
	\WisoDS(\eta)
	\equiv 
	\Delta W\left[(\eta,\xpw_{1})
	\rightarrow
	(\eta_{0},\xpw_{1})\right].
\end{equation}
\par
Regarding Eq.~\eqref{app_eq:def_DeltaG_DS}, 
let $\Phi_{\alpha \beta}(\bm{\Gamma},\eta)$ be the sum of the potential 
energy between the different fluids in Eq.~\eqref{eq:lj_potential_eta}:
\begin{equation}
	\Phi_{\alpha \beta}(\bm{\Gamma},\eta,\xpw_{1}) 
	= 
	\sum_{i\in\alpha}\sum_{j\in\beta}
	\phi_{\alpha\beta}(r_{ij})
	=
	\sum_{i\in\alpha}\sum_{j\in\beta}
	\eta \phi_{\alpha\alpha}{\left(r_{ij}\right)},
\end{equation}
then, the integrand 
$\left<\frac{\partial \Hamil(\eta',\xpw_{1})}{\partial \eta'}\right>$ 
in Eq.~\eqref{app_eq:def_DeltaG_DS} is written by
\begin{align}
	\nonumber
	\left<
	\frac{\partial \Hamil(\eta',\xpw_{1})
	}{
		\partial \eta'}
	\right>
	&=
	\left<
	\frac{\ptl \Phi_{\alpha\beta}(\eta',\xpw_{1}) }{\ptl \eta'}
	\right>
	=
	\left<
	\frac{\ptl}{\ptl \eta'}
	\sum_{i\in\alpha}^{N_{\alpha}}
	\sum_{j\in\beta}^{N_{\beta}}
	\phi_{\alpha\beta}(r_{ij})
	\right>
	\\ \nonumber
	&=
	\left<
	\sum_{i\in\alpha}^{N_{\alpha}}
	\sum_{j\in\beta}^{N_{\beta}}
	\phi_{\alpha\alpha}(r_{ij})
	\right>
	=\frac{1}{\eta'}
	\left<
	\sum_{i\in\alpha}^{N_{\alpha}}
	\sum_{j\in\beta}^{N_{\beta}}
	\phi_{\alpha\beta}(r_{ij})
	\right>
	\\
	&=
	\frac{\angb{
			\Phi_{\alpha\beta}(\eta',\xpw_{1})}
	}{\eta'}, 
	\label{app_eq:ptl_H_ptl_eta}
\end{align}
where $\left<\Phi_{\alpha\beta}(\eta',\xpw_{1})\right>$ can be 
easily obtained in the MD system.
By inserting Eq.~\eqref{app_eq:ptl_H_ptl_eta} into 
Eq.~\eqref{app_eq:def_DeltaG_DS} and further 
inserting it into Eq.~\eqref{app_eq:def_Wsys_DS}, 
the work for isolation results in
\begin{align}
	\WisoDS(\eta)
	=
	\int_{\eta_{0}}^{\eta} 
	\frac{1}{ \eta'}
	\left[-\frac{ \angb{\Phi_{\alpha\beta}(\eta',\xpw_{1})} }{A}
	\right]
	\mathrm{d}\eta'
	-
	\Wpc\left[(\eta,\xpw_{1})
	\rightarrow
	(\eta_{0},\xpw_{1})\right]
	\label{app_eq:w_iso_DS},
\end{align}
where $-\frac{\angb{\Phi_{\alpha\beta}(\eta',\xpw_{1})}}{A}$
is the average LL potential energy per area
with its sign reverted.
\par
In this study, we prepared multiple equilibrium systems 
with different miscibility parameter $\eta\in[0.01,1]$, 
and calculated the time average 
of the average LL potential energy over 20~ns 
for each equilibrium system to numerically integrate
the 1st term of the RHS of Eq.~\eqref{app_eq:w_iso_DS}. 
\par
On the other hand, in the PW method, we used the PW 
position $\xpw$ as the coupling parameter, and 
reproduce the reference system by pushing the PWs 
as in Fig.~\ref{fig:system}~(b-ii), 
\ie decreasing $\xpw$ from $\xpw_{1}$ down to 
$\xpw_{0}$ so that the two liquids were completely 
isolated. For this work, we define the work for isolation by the extended-PW denoted by $\WisoPW$ given by
\begin{equation}
	\WisoPW(\eta)
	\equiv
	\Delta W\left[(\eta,\xpw_{1})
	\rightarrow(\eta,\xpw_{0})\right].
\end{equation}
In this case, the difference of $G$ is written by
\begin{align}
	G(\eta,\xpw_{1})- G(\eta,\xpw_{0})
	=
	\int_{\xpw_{0}}^{\xpw_{1}} 
	\angb{
		\frac{\partial \Hamil(\eta,\xpwp)}{\partial \xpwp} 
	} \mrd \xpwp,
	\label{app_eq:def_DeltaG_PW}
\end{align}
It is obvious from the PW-fluid potential 
functions in Eqs.~\eqref{eq:potential_pwa}, 
\eqref{eq:potential_pwb} and \eqref{eq:xpwb=-xpwa} 
that the Hamiltonian derivative is reduced to 
the forces on the PWs as 
\begin{align}
	\nonumber
	\left<
	\frac{\partial \Hamil(\eta, \xpw)
	}{
		\partial \xpw}
	\right>
	&=
	\angb{
		-F_{\mathrm{fluid-pw}\alpha}(\eta, \xpw)
		+F_{\mathrm{fluid-pw}\beta}(\eta, \xpw)
	}
	\\ 
	&=
	-\angb{
		F_{\mathrm{fluid-pw}\alpha}(\eta, \xpw)
	}
	+\angb{
		F_{\mathrm{fluid-pw}\beta}(\eta, \xpw)
	},
	\label{app_eq:ptl_H_ptl_xpw}
\end{align}
where the average forces from the liquid $\alpha$ 
on PW$\alpha$ (right) and $\beta$ on PW$\beta$ (left) 
are defined by 
\begin{equation}
	\angb{F_{\mathrm{fluid-pw}\alpha}(\eta, \xpw)}
	=
	\angb{-
		\sum_{i\in \alpha} 
		\frac{\ptl \phi_{\mathrm{fluid-pw}\alpha}(x^{\prime}_{i})}{\ptl \xpwa}
	},
	\label{app_eq:force_on_PWa}
\end{equation}
and
\begin{equation}
	\angb{F_{\mathrm{fluid-pw}\beta}(\eta, \xpw)}
	=
	\angb{
		\sum_{i\in \beta} 
		\frac{\ptl \phi_{\mathrm{fluid-pw}\beta}(x^{\prime}_{i})}{\ptl \xpwb}
	},
	\label{app_eq:force_on_PWb}
\end{equation}
respectively. 
By using Eqs.~\eqref{app_eq:ptl_H_ptl_xpw} and \eqref{app_eq:def_Wsys_DS}, 
the work for isolation for the present systems is obtained by
\begin{align}
	\nonumber
	\WisoPW(\eta)
	&=
	-\frac{G(\eta,\xpw_{1})- G(\eta,\xpw_{0})}{A} 
	- 
	\Wpc\left[(\eta,\xpw_{1})
	\rightarrow
	(\eta,\xpw_{0})\right]
	\\
	\nonumber
	&=
	-\int_{\xpw_{1}}^{\xpw_{0}} 
	\frac{\angb{F_{\mathrm{fluid-pw}\alpha}(\eta, \xpwp)}}{A}
	\mrd \xpwp
	+
	\int_{\xpw_{1}}^{\xpw_{0}} 
	\frac{\angb{F_{\mathrm{fluid-pw}\beta}(\eta, \xpwp)}}{A}
	\mrd \xpwp
	\\
	&-
	\Wpc\left[(\eta,\xpw_{1})
	\rightarrow
	(\eta,\xpw_{0})\right]
	\label{app_eq:w_iso_PW}.
\end{align}
Note $\xpw_{1}>\xpw_{0}$, and also note that 
\begin{equation}
	\frac{
		\angb{F_{\mathrm{fluid-pw}\alpha}}
	}{A}
	\approx
	-\frac{\angb{F_{\mathrm{fluid-pw}\beta}}
	}{A}
	\geq 0
\end{equation}
holds for the forces from the property of 
$\phi_\mathrm{fluid-pw}$ and symmetry.
\section{Work for isolation by the extended-DS method}
\label{appsec:DS_result}
\begin{figure}[t]
	\begin{center}
		\vspace{5mm}
		\includegraphics[width=0.7\linewidth]{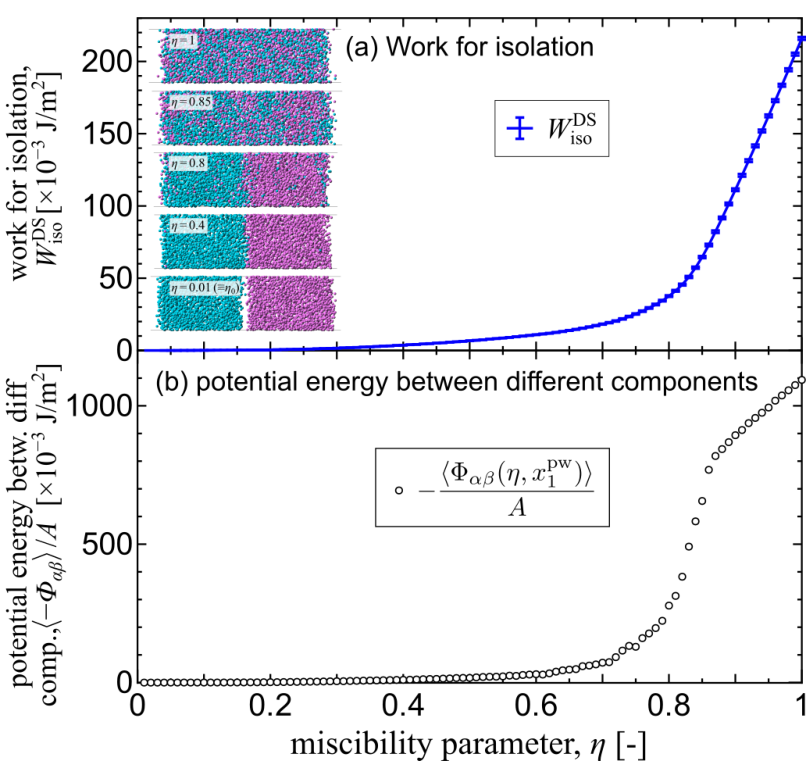}
	\end{center}
	\caption{
		\label{fig:result_DS}		
		(a) Work for isolation $\WisoDS$ obtained by the DS method in Eq.~\eqref{app_eq:w_iso_DS}, 
		and (b) average potential  energy between different components 
		$\alpha$ and $\beta$ for various miscibility parameter $\eta$. The inset corresponds to Fig.~\ref{fig:system}~(a-i).
		The error bars for (b) were smaller than the size of the symbol.
	}
\end{figure}
Figure~\ref{fig:result_DS}~(a) shows the work for isolation 
$\WisoDS(\eta)$ 
obtained by the DS method in 
Eq.~\eqref{app_eq:w_iso_DS} 
as a function of the miscibility parameter $\eta$, where the average potential energy between $\alpha$ and $\beta$ per area 
$-\frac{\angb{\Phi_{\alpha\beta}(\eta,\xpw_{1})}}{A}$ 
in Fig.~\ref{fig:result_DS}~(b) was used.
Note that the PWs did not interact with the liquids at 
the PW position $\xpw=\xpw_{1}$, and the PWs are not 
shown in the inset.
The potential energy $\Phi_{\alpha\beta}$ was always 
negative and was almost zero for small $\eta$ values 
because the two liquids were not mixed as observed in
the enlarged snapshot at $\eta=0.4$ in Fig.~\ref{fig:stress_dis}.
Hence, the resulting $\Wiso$ monotonically increased 
with a small gradient up to about $\eta=0.4$. 
For $\eta>0.5$,  $-\frac{\angb{\Phi_{\alpha\beta}}}{A}$ 
considerably increased, and the increase became
steep especially above around $\eta=0.7$, and the 
resulting $\WisoDS(\eta)$ also showed large increase
around $0.7 < \eta < 0.85$. 
Finally above around $\eta=0.85$, the steep increase 
of $-\frac{\angb{\Phi_{\alpha\beta}}}{A}$ 
calmed down. This turning point value of $\eta=0.85$ 
matched the point above which $\gab$ became zero 
in Fig.~\ref{fig:stress_dis}.
%
%

\begin{thebibliography}{34}%
	\makeatletter
	\providecommand \@ifxundefined [1]{%
		\@ifx{#1\undefined}
	}%
	\providecommand \@ifnum [1]{%
		\ifnum #1\expandafter \@firstoftwo
		\else \expandafter \@secondoftwo
		\fi
	}%
	\providecommand \@ifx [1]{%
		\ifx #1\expandafter \@firstoftwo
		\else \expandafter \@secondoftwo
		\fi
	}%
	\providecommand \natexlab [1]{#1}%
	\providecommand \enquote  [1]{``#1''}%
	\providecommand \bibnamefont  [1]{#1}%
	\providecommand \bibfnamefont [1]{#1}%
	\providecommand \citenamefont [1]{#1}%
	\providecommand \href@noop [0]{\@secondoftwo}%
	\providecommand \href [0]{\begingroup \@sanitize@url \@href}%
	\providecommand \@href[1]{\@@startlink{#1}\@@href}%
	\providecommand \@@href[1]{\endgroup#1\@@endlink}%
	\providecommand \@sanitize@url [0]{\catcode `\\12\catcode `\$12\catcode
		`\&12\catcode `\#12\catcode `\^12\catcode `\_12\catcode `\%12\relax}%
	\providecommand \@@startlink[1]{}%
	\providecommand \@@endlink[0]{}%
	\providecommand \url  [0]{\begingroup\@sanitize@url \@url }%
	\providecommand \@url [1]{\endgroup\@href {#1}{\urlprefix }}%
	\providecommand \urlprefix  [0]{URL }%
	\providecommand \Eprint [0]{\href }%
	\providecommand \doibase [0]{https://doi.org/}%
	\providecommand \selectlanguage [0]{\@gobble}%
	\providecommand \bibinfo  [0]{\@secondoftwo}%
	\providecommand \bibfield  [0]{\@secondoftwo}%
	\providecommand \translation [1]{[#1]}%
	\providecommand \BibitemOpen [0]{}%
	\providecommand \bibitemStop [0]{}%
	\providecommand \bibitemNoStop [0]{.\EOS\space}%
	\providecommand \EOS [0]{\spacefactor3000\relax}%
	\providecommand \BibitemShut  [1]{\csname bibitem#1\endcsname}%
	\let\auto@bib@innerbib\@empty
	\bibitem [{\citenamefont {Kirkwood}(1935)}]{kirkwood_statistical_1935}%
	\BibitemOpen
	\bibfield  {author} {\bibinfo {author} {\bibfnamefont {J.~G.}\ \bibnamefont
			{Kirkwood}},\ }\href {https://doi.org/10.1063/1.1749657} {\bibfield
		{journal} {\bibinfo  {journal} {J. Chem. Phys.}\ }\textbf {\bibinfo {volume}
			{3}},\ \bibinfo {pages} {300} (\bibinfo {year} {1935})}\BibitemShut {NoStop}%
	\bibitem [{\citenamefont {Kirkwood}\ and\ \citenamefont
		{Buff}(1951)}]{kirkwood_statistical_1951}%
	\BibitemOpen
	\bibfield  {author} {\bibinfo {author} {\bibfnamefont {J.~G.}\ \bibnamefont
			{Kirkwood}}\ and\ \bibinfo {author} {\bibfnamefont {F.~P.}\ \bibnamefont
			{Buff}},\ }\href {https://doi.org/10.1063/1.1748352} {\bibfield  {journal}
		{\bibinfo  {journal} {J. Chem. Phys.}\ }\textbf {\bibinfo {volume} {19}},\
		\bibinfo {pages} {774} (\bibinfo {year} {1951})}\BibitemShut {NoStop}%
	\bibitem [{\citenamefont {Kirkwood}\ and\ \citenamefont
		{Buff}(1949)}]{Kirkwood1949}%
	\BibitemOpen
	\bibfield  {author} {\bibinfo {author} {\bibfnamefont {J.~G.}\ \bibnamefont
			{Kirkwood}}\ and\ \bibinfo {author} {\bibfnamefont {F.~P.}\ \bibnamefont
			{Buff}},\ }\href {https://doi.org/10.1063/1.1747248} {\bibfield  {journal}
		{\bibinfo  {journal} {J. Chem. Phys.}\ }\textbf {\bibinfo {volume} {17}},\
		\bibinfo {pages} {338} (\bibinfo {year} {1949})}\BibitemShut {NoStop}%
	\bibitem [{\citenamefont {Bakker}(1928)}]{Bakker1928}%
	\BibitemOpen
	\bibfield  {author} {\bibinfo {author} {\bibfnamefont {G.}~\bibnamefont
			{Bakker}},\ }\href@noop {} {\emph {\bibinfo {title} {Kapillarit{\"a}t und
				Oberfl{\"a}chenspannung}}},\ Vol.~\bibinfo {volume} {6}\ (\bibinfo
	{publisher} {Wien-Harms},\ \bibinfo {year} {1928})\BibitemShut {NoStop}%
	\bibitem [{\citenamefont {Allen}\ and\ \citenamefont
		{Tildesley}(1987)}]{Allen1989}%
	\BibitemOpen
	\bibfield  {author} {\bibinfo {author} {\bibfnamefont {M.~P.}\ \bibnamefont
			{Allen}}\ and\ \bibinfo {author} {\bibfnamefont {D.~J.}\ \bibnamefont
			{Tildesley}},\ }\href@noop {} {\emph {\bibinfo {title} {Computer Simulation
				of Liquids}}}\ (\bibinfo  {publisher} {Oxford University Press},\ \bibinfo
	{year} {1987})\BibitemShut {NoStop}%
	\bibitem [{\citenamefont {Todd}\ \emph {et~al.}(1995)\citenamefont {Todd},
		\citenamefont {Evans},\ and\ \citenamefont {Daivis}}]{Todd1995}%
	\BibitemOpen
	\bibfield  {author} {\bibinfo {author} {\bibfnamefont {B.~D.}\ \bibnamefont
			{Todd}}, \bibinfo {author} {\bibfnamefont {D.~J.}\ \bibnamefont {Evans}},\
		and\ \bibinfo {author} {\bibfnamefont {P.~J.}\ \bibnamefont {Daivis}},\
	}\href {https://doi.org/10.1103/PhysRevE.52.1627} {\bibfield  {journal}
		{\bibinfo  {journal} {Phys. Rev. E}\ }\textbf {\bibinfo {volume} {52}},\
		\bibinfo {pages} {1627} (\bibinfo {year} {1995})}\BibitemShut {NoStop}%
	\bibitem [{\citenamefont {Shi}\ \emph {et~al.}(2023)\citenamefont {Shi},
		\citenamefont {Smith}, \citenamefont {Santiso},\ and\ \citenamefont
		{Gubbins}}]{Shi2023}%
	\BibitemOpen
	\bibfield  {author} {\bibinfo {author} {\bibfnamefont {K.}~\bibnamefont
			{Shi}}, \bibinfo {author} {\bibfnamefont {E.~R.}\ \bibnamefont {Smith}},
		\bibinfo {author} {\bibfnamefont {E.~E.}\ \bibnamefont {Santiso}},\ and\
		\bibinfo {author} {\bibfnamefont {K.~E.}\ \bibnamefont {Gubbins}},\ }\href
	{https://doi.org/10.1063/5.0132487} {\bibfield  {journal} {\bibinfo
			{journal} {J. Chem. Phys.}\ }\textbf {\bibinfo {volume} {158}},\ \bibinfo
		{pages} {040901} (\bibinfo {year} {2023})}\BibitemShut {NoStop}%
	\bibitem [{\citenamefont {Nishida}\ \emph {et~al.}(2014)\citenamefont
		{Nishida}, \citenamefont {Surblys}, \citenamefont {Yamaguchi}, \citenamefont
		{Kuroda}, \citenamefont {Kagawa}, \citenamefont {Nakajima},\ and\
		\citenamefont {Fujimura}}]{Nishida2014}%
	\BibitemOpen
	\bibfield  {author} {\bibinfo {author} {\bibfnamefont {S.}~\bibnamefont
			{Nishida}}, \bibinfo {author} {\bibfnamefont {D.}~\bibnamefont {Surblys}},
		\bibinfo {author} {\bibfnamefont {Y.}~\bibnamefont {Yamaguchi}}, \bibinfo
		{author} {\bibfnamefont {K.}~\bibnamefont {Kuroda}}, \bibinfo {author}
		{\bibfnamefont {M.}~\bibnamefont {Kagawa}}, \bibinfo {author} {\bibfnamefont
			{T.}~\bibnamefont {Nakajima}},\ and\ \bibinfo {author} {\bibfnamefont
			{H.}~\bibnamefont {Fujimura}},\ }\href
	{https://doi.org/http://dx.doi.org/10.1063/1.4865254} {\bibfield  {journal}
		{\bibinfo  {journal} {J. Chem. Phys.}\ }\textbf {\bibinfo {volume} {140}},\
		\bibinfo {pages} {074707} (\bibinfo {year} {2014})}\BibitemShut {NoStop}%
	\bibitem [{\citenamefont {Yamaguchi}\ \emph {et~al.}(2019)\citenamefont
		{Yamaguchi}, \citenamefont {Kusudo}, \citenamefont {Surblys}, \citenamefont
		{Omori},\ and\ \citenamefont {Kikugawa}}]{Yamaguchi2019}%
	\BibitemOpen
	\bibfield  {author} {\bibinfo {author} {\bibfnamefont {Y.}~\bibnamefont
			{Yamaguchi}}, \bibinfo {author} {\bibfnamefont {H.}~\bibnamefont {Kusudo}},
		\bibinfo {author} {\bibfnamefont {D.}~\bibnamefont {Surblys}}, \bibinfo
		{author} {\bibfnamefont {T.}~\bibnamefont {Omori}},\ and\ \bibinfo {author}
		{\bibfnamefont {G.}~\bibnamefont {Kikugawa}},\ }\href
	{https://doi.org/10.1063/1.5053881} {\bibfield  {journal} {\bibinfo
			{journal} {J. Chem. Phys.}\ }\textbf {\bibinfo {volume} {150}},\ \bibinfo
		{pages} {044701} (\bibinfo {year} {2019})}\BibitemShut {NoStop}%
	\bibitem [{\citenamefont {Hayoun}\ \emph {et~al.}(1988)\citenamefont {Hayoun},
		\citenamefont {Meyer}, \citenamefont {Mareschal}, \citenamefont {Ciccotti},\
		and\ \citenamefont {Turq}}]{Hayoun1988}%
	\BibitemOpen
	\bibfield  {author} {\bibinfo {author} {\bibfnamefont {M.}~\bibnamefont
			{Hayoun}}, \bibinfo {author} {\bibfnamefont {M.}~\bibnamefont {Meyer}},
		\bibinfo {author} {\bibfnamefont {M.}~\bibnamefont {Mareschal}}, \bibinfo
		{author} {\bibfnamefont {G.}~\bibnamefont {Ciccotti}},\ and\ \bibinfo
		{author} {\bibfnamefont {P.}~\bibnamefont {Turq}},\ }\bibinfo {title}
	{Molecular dynamics simulation of a liquid-liquid interface},\ in\ \href
	{https://doi.org/10.1007/978-1-4613-1023-5_24} {\emph {\bibinfo {booktitle}
			{Chemical Reactivity in Liquids: Fundamental Aspects}}},\ \bibinfo {editor}
	{edited by\ \bibinfo {editor} {\bibfnamefont {M.}~\bibnamefont {Moreau}}\
		and\ \bibinfo {editor} {\bibfnamefont {P.}~\bibnamefont {Turq}}}\ (\bibinfo
	{publisher} {Springer US},\ \bibinfo {address} {Boston, MA},\ \bibinfo {year}
	{1988})\ pp.\ \bibinfo {pages} {279--286}\BibitemShut {NoStop}%
	\bibitem [{\citenamefont {Benjamin}(1997)}]{benjamin_molecular_1997}%
	\BibitemOpen
	\bibfield  {author} {\bibinfo {author} {\bibfnamefont {I.}~\bibnamefont
			{Benjamin}},\ }\href {https://doi.org/10.1146/annurev.physchem.48.1.407}
	{\bibfield  {journal} {\bibinfo  {journal} {Ann. Rev. Phys. Chem.}\ }\textbf
		{\bibinfo {volume} {48}},\ \bibinfo {pages} {407} (\bibinfo {year}
		{1997})}\BibitemShut {NoStop}%
	\bibitem [{\citenamefont {Feria}\ \emph {et~al.}(2022)\citenamefont {Feria},
		\citenamefont {Algaba}, \citenamefont {Míguez}, \citenamefont {Mejía},\
		and\ \citenamefont {Blas}}]{feria_molecular_2022}%
	\BibitemOpen
	\bibfield  {author} {\bibinfo {author} {\bibfnamefont {E.}~\bibnamefont
			{Feria}}, \bibinfo {author} {\bibfnamefont {J.}~\bibnamefont {Algaba}},
		\bibinfo {author} {\bibfnamefont {J.~M.}\ \bibnamefont {Míguez}}, \bibinfo
		{author} {\bibfnamefont {A.}~\bibnamefont {Mejía}},\ and\ \bibinfo {author}
		{\bibfnamefont {F.~J.}\ \bibnamefont {Blas}},\ }\href
	{https://doi.org/10.1039/D1CP05346A} {\bibfield  {journal} {\bibinfo
			{journal} {Phys. Chem. Chem. Phys.}\ }\textbf {\bibinfo {volume} {24}},\
		\bibinfo {pages} {5371} (\bibinfo {year} {2022})}\BibitemShut {NoStop}%
	\bibitem [{\citenamefont {Sega}\ \emph {et~al.}(2016)\citenamefont {Sega},
		\citenamefont {F{\'a}bi{\'a}n}, \citenamefont {Horvai},\ and\ \citenamefont
		{Jedlovszky}}]{Sega2016}%
	\BibitemOpen
	\bibfield  {author} {\bibinfo {author} {\bibfnamefont {M.}~\bibnamefont
			{Sega}}, \bibinfo {author} {\bibfnamefont {B.}~\bibnamefont
			{F{\'a}bi{\'a}n}}, \bibinfo {author} {\bibfnamefont {G.}~\bibnamefont
			{Horvai}},\ and\ \bibinfo {author} {\bibfnamefont {P.}~\bibnamefont
			{Jedlovszky}},\ }\href {https://doi.org/10.1021/acs.jpcc.6b09880} {\bibfield
		{journal} {\bibinfo  {journal} {J. Phys. Chem. C}\ }\textbf {\bibinfo
			{volume} {120}},\ \bibinfo {pages} {27468} (\bibinfo {year}
		{2016})}\BibitemShut {NoStop}%
	\bibitem [{\citenamefont {Hantal}\ \emph {et~al.}(2020)\citenamefont {Hantal},
		\citenamefont {F{\'a}bi{\'a}n}, \citenamefont {Sega},\ and\ \citenamefont
		{Jedlovszky}}]{Hantal2020}%
	\BibitemOpen
	\bibfield  {author} {\bibinfo {author} {\bibfnamefont {G.}~\bibnamefont
			{Hantal}}, \bibinfo {author} {\bibfnamefont {B.}~\bibnamefont
			{F{\'a}bi{\'a}n}}, \bibinfo {author} {\bibfnamefont {M.}~\bibnamefont
			{Sega}},\ and\ \bibinfo {author} {\bibfnamefont {P.}~\bibnamefont
			{Jedlovszky}},\ }\href {https://doi.org/10.1016/j.molliq.2020.112872}
	{\bibfield  {journal} {\bibinfo  {journal} {J. Mol. Liquids}\ }\textbf
		{\bibinfo {volume} {306}},\ \bibinfo {pages} {112872} (\bibinfo {year}
		{2020})}\BibitemShut {NoStop}%
	\bibitem [{\citenamefont {Leroy}\ \emph {et~al.}(2009)\citenamefont {Leroy},
		\citenamefont {Dos~Santos},\ and\ \citenamefont
		{M{\"u}ller-Plathe}}]{Leroy2009}%
	\BibitemOpen
	\bibfield  {author} {\bibinfo {author} {\bibfnamefont {F.}~\bibnamefont
			{Leroy}}, \bibinfo {author} {\bibfnamefont {D.~J. V.~A.}\ \bibnamefont
			{Dos~Santos}},\ and\ \bibinfo {author} {\bibfnamefont {F.}~\bibnamefont
			{M{\"u}ller-Plathe}},\ }\href {https://doi.org/10.1002/marc.200800746}
	{\bibfield  {journal} {\bibinfo  {journal} {Macromol. Rapid Commun.}\
		}\textbf {\bibinfo {volume} {30}},\ \bibinfo {pages} {864} (\bibinfo {year}
		{2009})}\BibitemShut {NoStop}%
	\bibitem [{\citenamefont {Leroy}\ and\ \citenamefont
		{M{\"u}ller-Plathe}(2015)}]{Leroy2015}%
	\BibitemOpen
	\bibfield  {author} {\bibinfo {author} {\bibfnamefont {F.}~\bibnamefont
			{Leroy}}\ and\ \bibinfo {author} {\bibfnamefont {F.}~\bibnamefont
			{M{\"u}ller-Plathe}},\ }\href {https://doi.org/10.1021/acs.langmuir.5b01394}
	{\bibfield  {journal} {\bibinfo  {journal} {Langmuir}\ }\textbf {\bibinfo
			{volume} {31}},\ \bibinfo {pages} {8335} (\bibinfo {year}
		{2015})}\BibitemShut {NoStop}%
	\bibitem [{\citenamefont {Kandu{\v{c}}}(2017)}]{Kanduc2017}%
	\BibitemOpen
	\bibfield  {author} {\bibinfo {author} {\bibfnamefont {M.}~\bibnamefont
			{Kandu{\v{c}}}},\ }\href {https://dx.doi.org/10.1063/1.4990741} {\bibfield
		{journal} {\bibinfo  {journal} {J. Chem. Phys.}\ }\textbf {\bibinfo {volume}
			{147}},\ \bibinfo {pages} {174701} (\bibinfo {year} {2017})}\BibitemShut
	{NoStop}%
	\bibitem [{\citenamefont {Russo}\ \emph {et~al.}(2019)\citenamefont {Russo},
		\citenamefont {{Dur{\'a}n-Olivencia}}, \citenamefont {Kalliadasis},\ and\
		\citenamefont {Hartkamp}}]{Russo2019}%
	\BibitemOpen
	\bibfield  {author} {\bibinfo {author} {\bibfnamefont {A.}~\bibnamefont
			{Russo}}, \bibinfo {author} {\bibfnamefont {M.~A.}\ \bibnamefont
			{{Dur{\'a}n-Olivencia}}}, \bibinfo {author} {\bibfnamefont {S.}~\bibnamefont
			{Kalliadasis}},\ and\ \bibinfo {author} {\bibfnamefont {R.}~\bibnamefont
			{Hartkamp}},\ }\href {https://doi.org/10.1063/1.5094911} {\bibfield
		{journal} {\bibinfo  {journal} {J Chem. Phys.}\ }\textbf {\bibinfo {volume}
			{150}},\ \bibinfo {pages} {214705} (\bibinfo {year} {2019})}\BibitemShut
	{NoStop}%
	\bibitem [{\citenamefont {Surblys}\ \emph {et~al.}(2018)\citenamefont
		{Surblys}, \citenamefont {Leroy}, \citenamefont {Yamaguchi},\ and\
		\citenamefont {M{\"u}ller-Plathe}}]{Surblys2018}%
	\BibitemOpen
	\bibfield  {author} {\bibinfo {author} {\bibfnamefont {D.}~\bibnamefont
			{Surblys}}, \bibinfo {author} {\bibfnamefont {F.}~\bibnamefont {Leroy}},
		\bibinfo {author} {\bibfnamefont {Y.}~\bibnamefont {Yamaguchi}},\ and\
		\bibinfo {author} {\bibfnamefont {F.}~\bibnamefont {M{\"u}ller-Plathe}},\
	}\href {https://doi.org/10.1063/1.3601055} {\bibfield  {journal} {\bibinfo
			{journal} {J. Chem. Phys.}\ }\textbf {\bibinfo {volume} {148}},\ \bibinfo
		{eid} {134707} (\bibinfo {year} {2018})}\BibitemShut {NoStop}%
	\bibitem [{\citenamefont {Bistafa}\ \emph {et~al.}(2021)\citenamefont
		{Bistafa}, \citenamefont {Surblys}, \citenamefont {Kusudo},\ and\
		\citenamefont {Yamaguchi}}]{Bistafa2021}%
	\BibitemOpen
	\bibfield  {author} {\bibinfo {author} {\bibfnamefont {C.}~\bibnamefont
			{Bistafa}}, \bibinfo {author} {\bibfnamefont {D.}~\bibnamefont {Surblys}},
		\bibinfo {author} {\bibfnamefont {H.}~\bibnamefont {Kusudo}},\ and\ \bibinfo
		{author} {\bibfnamefont {Y.}~\bibnamefont {Yamaguchi}},\ }\href
	{https://doi.org/10.1063/5.0056718} {\bibfield  {journal} {\bibinfo
			{journal} {J. Chem. Phys.}\ }\textbf {\bibinfo {volume} {155}},\ \bibinfo
		{pages} {064703} (\bibinfo {year} {2021})}\BibitemShut {NoStop}%
	\bibitem [{\citenamefont {Shintaku}\ \emph {et~al.}(2024)\citenamefont
		{Shintaku}, \citenamefont {Oga}, \citenamefont {Yamaguchi}, \citenamefont
		{Kusudo}, \citenamefont {Smith},\ and\ \citenamefont {Omori}}]{Shintaku2024}%
	\BibitemOpen
	\bibfield  {author} {\bibinfo {author} {\bibfnamefont {M.}~\bibnamefont
			{Shintaku}}, \bibinfo {author} {\bibfnamefont {H.}~\bibnamefont {Oga}},
		\bibinfo {author} {\bibfnamefont {Y.}~\bibnamefont {Yamaguchi}}, \bibinfo
		{author} {\bibfnamefont {H.}~\bibnamefont {Kusudo}}, \bibinfo {author}
		{\bibfnamefont {E.~R.}\ \bibnamefont {Smith}},\ and\ \bibinfo {author}
		{\bibfnamefont {T.}~\bibnamefont {Omori}},\ }\href
	{https://doi.org/10.1063/5.0201973} {\bibfield  {journal} {\bibinfo
			{journal} {J. Chem. Phys.}\ }\textbf {\bibinfo {volume} {160}},\ \bibinfo
		{pages} {224502} (\bibinfo {year} {2024})}\BibitemShut {NoStop}%
	\bibitem [{\citenamefont {Kusudo}\ \emph {et~al.}(2021)\citenamefont {Kusudo},
		\citenamefont {Omori},\ and\ \citenamefont {Yamaguchi}}]{Kusudo2021}%
	\BibitemOpen
	\bibfield  {author} {\bibinfo {author} {\bibfnamefont {H.}~\bibnamefont
			{Kusudo}}, \bibinfo {author} {\bibfnamefont {T.}~\bibnamefont {Omori}},\ and\
		\bibinfo {author} {\bibfnamefont {Y.}~\bibnamefont {Yamaguchi}},\ }\href
	{https://doi.org/10.1063/5.0062889} {\bibfield  {journal} {\bibinfo
			{journal} {J. Chem. Phys.}\ }\textbf {\bibinfo {volume} {155}},\ \bibinfo
		{pages} {184103} (\bibinfo {year} {2021})}\BibitemShut {NoStop}%
	\bibitem [{\citenamefont {Kusudo}\ \emph {et~al.}(2023)\citenamefont {Kusudo},
		\citenamefont {Omori}, \citenamefont {Joly},\ and\ \citenamefont
		{Yamaguchi}}]{Kusudo2023}%
	\BibitemOpen
	\bibfield  {author} {\bibinfo {author} {\bibfnamefont {H.}~\bibnamefont
			{Kusudo}}, \bibinfo {author} {\bibfnamefont {T.}~\bibnamefont {Omori}},
		\bibinfo {author} {\bibfnamefont {L.}~\bibnamefont {Joly}},\ and\ \bibinfo
		{author} {\bibfnamefont {Y.}~\bibnamefont {Yamaguchi}},\ }\href
	{https://doi.org/10.1063/5.0171769} {\bibfield  {journal} {\bibinfo
			{journal} {J. Chem. Phys.}\ }\textbf {\bibinfo {volume} {159}},\ \bibinfo
		{pages} {161102} (\bibinfo {year} {2023})}\BibitemShut {NoStop}%
	\bibitem [{\citenamefont {Todd}\ and\ \citenamefont
		{Daivis}(2017)}]{Todd_Daivis_book}%
	\BibitemOpen
	\bibfield  {author} {\bibinfo {author} {\bibfnamefont {B.~D.}\ \bibnamefont
			{Todd}}\ and\ \bibinfo {author} {\bibfnamefont {P.~J.}\ \bibnamefont
			{Daivis}},\ }\href@noop {} {\emph {\bibinfo {title} {Nonequilibrium Molecular
				Dynamics: Theory, Algorithms and Applications}}}\ (\bibinfo  {publisher}
	{Cambridge University Press},\ \bibinfo {year} {2017})\BibitemShut {NoStop}%
	\bibitem [{\citenamefont {Leroy}\ and\ \citenamefont
		{M{\"u}ller-Plathe}(2010)}]{Leroy2010}%
	\BibitemOpen
	\bibfield  {author} {\bibinfo {author} {\bibfnamefont {F.}~\bibnamefont
			{Leroy}}\ and\ \bibinfo {author} {\bibfnamefont {F.}~\bibnamefont
			{M{\"u}ller-Plathe}},\ }\href {https://doi.org/10.1063/1.3458796} {\bibfield
		{journal} {\bibinfo  {journal} {J. Chem. Phys.}\ }\textbf {\bibinfo {volume}
			{133}},\ \bibinfo {pages} {044110} (\bibinfo {year} {2010})}\BibitemShut
	{NoStop}%
	\bibitem [{\citenamefont {Frenkel}\ and\ \citenamefont
		{Smit}(2001)}]{Frenkel2001}%
	\BibitemOpen
	\bibfield  {author} {\bibinfo {author} {\bibfnamefont {D.}~\bibnamefont
			{Frenkel}}\ and\ \bibinfo {author} {\bibfnamefont {B.}~\bibnamefont {Smit}},\
	}\href@noop {} {\emph {\bibinfo {title} {Understanding Molecular Simulation,
				Second Edition: From Algorithms to Applications (Computational Science
				Series, Vol. 1)}}}\ (\bibinfo  {publisher} {Academic Press},\ \bibinfo {year}
	{2001})\ pp.\ \bibinfo {pages} {168--172}\BibitemShut {NoStop}%
	\bibitem [{\citenamefont {Davidchack}\ and\ \citenamefont
		{Laird}(2003)}]{Davidchack2003}%
	\BibitemOpen
	\bibfield  {author} {\bibinfo {author} {\bibfnamefont {R.~L.}\ \bibnamefont
			{Davidchack}}\ and\ \bibinfo {author} {\bibfnamefont {B.~B.}\ \bibnamefont
			{Laird}},\ }\href {https://doi.org/10.1063/1.1563248} {\bibfield  {journal}
		{\bibinfo  {journal} {J. Chem. Phys.}\ }\textbf {\bibinfo {volume} {118}},\
		\bibinfo {pages} {7651} (\bibinfo {year} {2003})}\BibitemShut {NoStop}%
	\bibitem [{\citenamefont {di~Pasquale}\ and\ \citenamefont
		{Davidchack}(2022)}]{diPasquale2022}%
	\BibitemOpen
	\bibfield  {author} {\bibinfo {author} {\bibfnamefont {N.}~\bibnamefont
			{di~Pasquale}}\ and\ \bibinfo {author} {\bibfnamefont {R.~L.}\ \bibnamefont
			{Davidchack}},\ }\href {https://doi.org/10.1021/acs.jpca.2c00604} {\bibfield
		{journal} {\bibinfo  {journal} {J. Phys. Chem. A}\ }\textbf {\bibinfo
			{volume} {126}},\ \bibinfo {pages} {2134} (\bibinfo {year}
		{2022})}\BibitemShut {NoStop}%
	\bibitem [{\citenamefont {Surblys}\ \emph {et~al.}(2014)\citenamefont
		{Surblys}, \citenamefont {Yamaguchi}, \citenamefont {Kuroda}, \citenamefont
		{Kagawa}, \citenamefont {Nakajima},\ and\ \citenamefont
		{Fujimura}}]{Surblys2014}%
	\BibitemOpen
	\bibfield  {author} {\bibinfo {author} {\bibfnamefont {D.}~\bibnamefont
			{Surblys}}, \bibinfo {author} {\bibfnamefont {Y.}~\bibnamefont {Yamaguchi}},
		\bibinfo {author} {\bibfnamefont {K.}~\bibnamefont {Kuroda}}, \bibinfo
		{author} {\bibfnamefont {M.}~\bibnamefont {Kagawa}}, \bibinfo {author}
		{\bibfnamefont {T.}~\bibnamefont {Nakajima}},\ and\ \bibinfo {author}
		{\bibfnamefont {H.}~\bibnamefont {Fujimura}},\ }\href
	{https://doi.org/http://dx.doi.org/10.1063/1.4861039} {\bibfield  {journal}
		{\bibinfo  {journal} {J. Chem. Phys.}\ }\textbf {\bibinfo {volume} {140}},\
		\bibinfo {eid} {034505} (\bibinfo {year} {2014})}\BibitemShut {NoStop}%
	\bibitem [{\citenamefont {Kusudo}\ \emph {et~al.}(2019)\citenamefont {Kusudo},
		\citenamefont {Omori},\ and\ \citenamefont {Yamaguchi}}]{Kusudo2019}%
	\BibitemOpen
	\bibfield  {author} {\bibinfo {author} {\bibfnamefont {H.}~\bibnamefont
			{Kusudo}}, \bibinfo {author} {\bibfnamefont {T.}~\bibnamefont {Omori}},\ and\
		\bibinfo {author} {\bibfnamefont {Y.}~\bibnamefont {Yamaguchi}},\ }\href
	{https://doi.org/10.1063/1.5124014} {\bibfield  {journal} {\bibinfo
			{journal} {J. Chem. Phys.}\ }\textbf {\bibinfo {volume} {151}},\ \bibinfo
		{pages} {154501} (\bibinfo {year} {2019})}\BibitemShut {NoStop}%
	\bibitem [{\citenamefont {Rowlinson}\ and\ \citenamefont
		{Widom}(1982)}]{Rowlinson1982}%
	\BibitemOpen
	\bibfield  {author} {\bibinfo {author} {\bibfnamefont {J.~S.}\ \bibnamefont
			{Rowlinson}}\ and\ \bibinfo {author} {\bibfnamefont {B.}~\bibnamefont
			{Widom}},\ }\href@noop {} {\emph {\bibinfo {title} {Molecular Theory of
				Capillarity}}}\ (\bibinfo  {publisher} {Dover},\ \bibinfo {year} {1982})\
	p.~\bibinfo {pages} {45}\BibitemShut {NoStop}%
	\bibitem [{\citenamefont {Luo}\ and\ \citenamefont {Roux}(2010)}]{Luo2010}%
	\BibitemOpen
	\bibfield  {author} {\bibinfo {author} {\bibfnamefont {Y.}~\bibnamefont
			{Luo}}\ and\ \bibinfo {author} {\bibfnamefont {B.}~\bibnamefont {Roux}},\
	}\href {https://doi.org/10.1021/jz900079w} {\bibfield  {journal} {\bibinfo
			{journal} {J. Phys. Chem. Lett.}\ }\textbf {\bibinfo {volume} {1}},\ \bibinfo
		{pages} {183} (\bibinfo {year} {2010})}\BibitemShut {NoStop}%
	\bibitem [{\citenamefont {Bley}\ \emph {et~al.}(2017)\citenamefont {Bley},
		\citenamefont {Duvail}, \citenamefont {Guilbaud},\ and\ \citenamefont
		{Dufrêche}}]{Bley2017}%
	\BibitemOpen
	\bibfield  {author} {\bibinfo {author} {\bibfnamefont {M.}~\bibnamefont
			{Bley}}, \bibinfo {author} {\bibfnamefont {M.}~\bibnamefont {Duvail}},
		\bibinfo {author} {\bibfnamefont {P.}~\bibnamefont {Guilbaud}},\ and\
		\bibinfo {author} {\bibfnamefont {J.-F.}\ \bibnamefont {Dufrêche}},\ }\href
	{https://doi.org/10.1021/acs.jpcb.7b04011} {\bibfield  {journal} {\bibinfo
			{journal} {J. Phys. Chem. B}\ }\textbf {\bibinfo {volume} {121}},\ \bibinfo
		{pages} {9647} (\bibinfo {year} {2017})}\BibitemShut {NoStop}%
	\bibitem [{\citenamefont {Greiner}\ \emph {et~al.}(1995)\citenamefont
		{Greiner}, \citenamefont {Neise}, \citenamefont {St\"{o}cker},\ and\
		\citenamefont {Rischke}}]{Greiner1995}%
	\BibitemOpen
	\bibfield  {author} {\bibinfo {author} {\bibfnamefont {W.}~\bibnamefont
			{Greiner}}, \bibinfo {author} {\bibfnamefont {L.}~\bibnamefont {Neise}},
		\bibinfo {author} {\bibfnamefont {H.}~\bibnamefont {St\"{o}cker}},\ and\
		\bibinfo {author} {\bibfnamefont {D.}~\bibnamefont {Rischke}},\ }\href@noop
	{} {\emph {\bibinfo {title} {Thermodynamics and Statistical Mechanics
				(Classical Theoretical Physics)}}},\ \bibinfo {edition} {3rd}\ ed.\ (\bibinfo
	{publisher} {Springer},\ \bibinfo {year} {1995})\ pp.\ \bibinfo {pages}
	{43--48}\BibitemShut {NoStop}%
\end{thebibliography}
%
%
%
\end{document}